\documentclass[12pt,tightenlines,eqsecnum,floats,aps,amsmath,amssymb,superscriptaddress,nofootinbib,prd,showpacs]{revtex4}

\usepackage[dvips]{graphicx}
\usepackage[mathscr]{eucal}
\usepackage[latin1]{inputenc}
\usepackage{amsmath}
\usepackage{amsfonts}
\usepackage{amssymb}
\usepackage{pictex}
\usepackage{epsfig}
\usepackage{colordvi}
\usepackage{color}

\def\be{\nopagebreak[3]\begin{equation}}
\def\ee{\end{equation}}
\def\ba{\nopagebreak[3]\begin{eqnarray}}
\def\ea{\end{eqnarray}}
\def\bas{\nopagebreak[3]\begin{eqnarray*}}
\def\eas{\end{eqnarray*}}

\def\d{{\rm d}}
\def\w{\omega}

\newcommand{\teta}{\rlap{\lower2ex\hbox{$\,\tilde{}$}}\eta{}}

\newcommand{\bi}{\begin{itemize}}
\newcommand{\ei}{\end{itemize}}
\newcommand{\mc}[1]{\mathcal{#1}}

\def\lp{{\ell}_{\rm Pl}}

\def\q{{}^o\!q}
\def\e{{}^o\!e}
\def\w{{}^o\!\omega}

\newcommand{\p}{\partial}

\newcommand{\f}{\frac}

\usepackage{enumerate}

\usepackage{colordvi}
\newcounter{mnotecount}[section]

\def\f{\frac}

\begin{document}
\preprint{\vbox{\baselineskip=12pt \rightline{IGC-10/12-3}
}}

\title{Effective Dynamics in Bianchi Type II Loop Quantum Cosmology}
\author{Alejandro Corichi}\email{corichi@matmor.unam.mx}
\affiliation{Centro de Ciencias Matem\'aticas,
Universidad Nacional Aut\'onoma de M\'exico,
UNAM-Campus Morelia, A. Postal 61-3, Morelia, Michoac\'an 58090,
Mexico}
\affiliation{Center for Fundamental Theory, Institute for Gravitation and the Cosmos,
Pennsylvania State University, University Park
PA 16802, USA}
\author{Edison Montoya}
\email{edison@matmor.unam.mx}
 \affiliation{Instituto de F\'{\i}sica y
Matem\'aticas,  Universidad Michoacana de San Nicol\'as de
Hidalgo, Morelia, Michoac\'an, Mexico}
\affiliation{Centro de Ciencias Matem\'aticas,
Universidad Nacional Aut\'onoma de M\'exico,
UNAM-Campus Morelia, A. Postal 61-3, Morelia, Michoac\'an 58090,
Mexico}

\begin{abstract}
We numerically  investigate the solutions to the effective equations of the  Bianchi II model within the
``improved" Loop Quantum Cosmology (LQC) dynamics. The matter source is a massless scalar field. 
We perform  a systematic study of the space of solutions, and focus on the behavior of several geometrical
observables. We show that the big-bang singularity is 
replaced by a bounce and the point-like singularities do not saturate the energy density bound. 
There are up to three directional bounces in the scale factors, one global bounce
in the expansion, the shear presents up to four local maxima and can be zero at the bounce. This allows for
solutions with density larger than the maximal density for the isotropic and Bianchi I cases. The 
asymptotic behavior is shown to behave like that of a Bianchi I model, and the effective solutions connect anisotropic solutions
even when the shear is zero at the bounce. All known facts of Bianchi I are reproduced. In the ``vacuum limit", solutions are such that almost all the dynamics is due to the anisotropies. Since Bianchi II plays
an important role in the Bianchi IX model and the Belinskii, Khalatnikov, Lifshitz (BKL) conjecture, our results
can provide an intuitive understanding of the behavior in the vicinity of general space-like singularities, when loop-geometric corrections are present.

\end{abstract}

\maketitle

\section{Introduction}
\label{sec:1}

Loop Quantum Gravity (LQG) \cite{lqg} has recently emerged as a strong candidate for a quantum theory of gravity.
One of the main motivations for such a theory is to provide a solution to the big-bang singularity. Even when the full theory has still little to say about the initial singularity, a symmetry reduced theory, namely Loop Quantum Cosmology (LQC) \cite{lqc,AA,AS}, 
has been extremely successful at providing precise answers to that question. The theory is constructed by applying the methods of LQG to a symmetry reduced sector of general relativity. As examples of these 
reduced configurations,  several authors have studied cosmological models with a massless scalar field and geometrically  isotropic 
\cite{aps0, aps, aps2, closed, open}, homogeneous and anisotropic \cite{bianchiold,bianchiI,bianchiII,bianchiIX}, and some inhomogeneous cosmologies \cite{gowdy}. 
The common theme among these models is that they resolve the big bang singularity. The way singularity resolution occurs is by means of physical observables (in the sense of Dirac) whose expectation values (or spectrum)  have been shown to be bounded \cite{slqc,tomasz,param,geometric}. These results benefit from uniqueness results that guaranty the consistency of the so called ``improved dynamics" \cite{unique,geometric}.

In a sense, isotropic LQC can be seen as a realization of one of the main objective of LQG, namely, 
to solve the big-bang singularity.  In this case the singularity is replaced by a bounce that occurs precisely when the matter density enters the Planck regime. At this energy density,
the quantum effects create a repulsive force, the would-be singularity is avoided and the resulting
`quantum' spacetime is larger than one might be led to believe. As has been shown in detail, when the density decreases, the state very quickly leaves this quantum regime, and the universe returns to being well described by general relativity \cite{lqc}. With the final goal in mind of investigating the most general issue of
singularity resolution within LQG, one expects to gain useful insights from results obtained for less symmetric models. The simplest such model is given by Bianchi I  
cosmologies \cite{bianchiI} where no spatial curvature exists. Even for this case we do not possess yet a full exact evolution of the quantum equations of motion. 

For both isotropic and anisotropic models, a very useful tool has been the use of the so-called effective description, a `classical theory' (in the sense that it does not contain $\hbar$) that has information about the geometric discreetness contained in loop quantum gravity. In isotropic models, the solutions 
to the effective equations 
have been shown to approximate the dynamics of semiclassical states in the full quantum theory with very good accuracy \cite{aps2,closed,open,CM}. In the case of anisotropic models we also expect that the effective solutions 
play an important role for describing  the evolution of semiclassical states. Thus, we shall adopt the viewpoint that it is justified to study the effective dynamics for those cosmological models, as a way to learn about the full `loopy' quantum dynamics.

So far, the only anisotropic model that has been explored in detail, through the so called {\it effective equations}, is the LQC Bianchi I model \cite{bianchiold}. It is then natural to explore this issue for more complicated anisotropic models.
In this paper we will study the effective equations obtained from the ``improved'' LQC dynamics of the Bianchi II model \cite{bianchiII}. The Bianchi II cosmological model represents the simplest case that posses spatial curvature, from which the Bianchi I model can be recovered when a parameter measuring the spatial curvature contribution is `switched off'. 
The Bianchi II model possesses another interesting feature, namely, it lies at the heart of the Belinskii, Khalatnikov, Lifshitz (BKL) conjecture \cite{BKL1,BKL2,BKL-AHS}, which suggest that, 
as one approaches space-like
singularities, the behavior of the system undergoes Bianchi I phases with Bianchi II transitions. 
One question that remains open though is whether this BKL behavior will survive in the effective theory.
That is, will the oscillations between Bianchi I phases occur far from the Planck scale? or will the loop-geometric effects prevent this `mixmaster' behavior to manifest itself?
From this point of view, it is important to study in detail the Bianchi II
solutions. 
The most natural strategy to gain this intuition is to perform a systematic study of  the solutions to the effective equations, under the assumption that they describe correctly the quantum dynamics.  But even if 
the effective solutions do not describe correctly the quantum dynamics of the semiclassical states in some regime, we need to study them in detail in order to compare them to the full quantum dynamics --when available-- and prove their validity.
One can also expect that this study will shed some light on the larger issue of understanding generic space-like singularities in LQG.

The purpose of this paper is to study in a systematic way the space of solutions of the effective equations. 
The matter source that we shall consider is a massless scalar field that plays the role of internal time. 
The objective is to understand the singularity resolution, the asymptotic behavior and the relation between the Bianchi II and Bianchi I models. The strategy will be to take limiting cases and compare them with known solutions. This will allow us to understand the new insights of the Bianchi II model. All the information will come from the set of observables that we define, namely, directional scale factors, Hubble parameters, expansion, matter density, density parameter, shear, shear parameter, Ricci scalar, curvature parameter and Kasner exponents.  With these tools in hand we shall compare the classical and the effective solutions. Later on, the isotropic limit will offer  interesting new insights into the Bianchi II dynamics, while the Bianchi I limit will allow us to confirm that our solutions agree with the previous results of \cite{bianchiold}. The symmetry reduction to the Locally Rotationally Symmetric (LRS) Bianchi II model will give us a way to find generic solutions with maximal density, at the bounce, larger than the critical density derived in the isotropic case \cite{aps2,slqc}. Further, the vacuum limit will allows us to study the extreme solutions where all the dynamical contributions come from the anisotropies. Finally, due to the fact that our work is numerical we will show that our solutions converge and evolve on the constraint surface.  The convergence is an important issue that needs to be shown, because one needs to ensure that the numerical methods are well implemented, and that the numerical solutions converge to the analytical solutions.

The structure of the paper is as follows. In Sec. II we recall the classical theory in metric and connection variables,
together with the basic observables to be studied. In Sec. III we introduce the effective theory and compute its equations of motion. 
Numerical solutions are explored in Sec. IV,  where
they are systematically studied taking the classical, isotropic, Bianchi I, maximal density and vacuum limits. 
We end with a discussion in Sec. V. There is one Appendix where we study the convergence and evolution
of the conserved quantities. Throughout the manuscript we assume units where $c=1$, the other constants are written explicitly.

\section{Classical Dynamics}
\label{sec:2}

The metric for a Bianchi II model can be written as
\be \label{metric} 
\d s^2= -N(\tau)^2 \d\tau^2 + a_1(\tau)^2\:(\d x-\alpha
z\:\d y)^2+a_2(\tau)^2\:\d y^2+a_3(\tau)^2\:\d z^2 \, ,
\ee
where the parameter $\alpha$ allows us to distinguish between Bianchi I  ($\alpha=0$) 
and Bianchi II ($\alpha=1$) cases. Bianchi I cosmological solutions are also interesting, since
they give information about the asymptotic behavior of Bianchi II. 
Classically, the Bianchi I case with a massless scalar field 
has solutions given by \cite{jacobs},
\be \label{metric_BI} 
\d s^2= - \d t^2 + t^{2k_1}\:\d x^2 +t^{2k_2}\:\d y^2+ t^{2k_3}\:\d z^2 
\ee
the so-called {\it Jacobs stiff perfect fluid solutions}.\footnote{The metric form \eqref{metric_BI} is taken from \cite{hsu_w}, where one can find a large class of analytical solutions to 
spatially homogeneous cosmological models.}
Here, the parameters $k_i$ are known as the {\it Kasner exponents} satisfying $k_1+k_2+k_3=\pm 1$ (the minus sign is inserted, for future reference, as we want to take into account the change of direction of
the expansion at the bounce) and $k_1^2+k_2^2+k_3^2+k_\phi^2= 1$. 
In the literature, a massless scalar field is also known as ``stiff matter'',\footnote{In the original article \cite{jacobs}, this solution is called ``Zel'dovich universe''. }
and satisfies the equation of state $P=\rho$,
where $P$ is pressure and $\rho$ is the energy density. All these solutions have an initial singularity at 
$t=0$, that can be of four types \cite{jacobs}:
\begin{enumerate}
\item {\it Point} type singularity. This means that $a_1,a_2,a_3 \rightarrow 0$ as $t\rightarrow 0$. This happen in the Jacobs solutions when $k_1,k_2,k_3 > 0$. 
\item {\it Cigar} type singularity. These occur when  $a_1,a_2 \rightarrow 0$ and $a_3 \rightarrow \infty$ as $t\rightarrow 0$. 
This happens when $k_1,k_2 > 0$ and $k_3<0$. (cyclic on 1,2,3) 
\item {\it Barrel} type singularity. Defined by $a_1,a_2 \rightarrow 0$ and $a_3$ approaching a finite value as $t\rightarrow 0$. 
This happen when $k_1,k_2 > 0$ and $k_3=0$. (cyclic on 1,2,3) 
\item {\it Pancake} type singularity. In this case, $a_1 \rightarrow 0$ and $a_2, a_3$ approaches a finite value as $t\rightarrow 0$ (cyclic on 1,2,3). This kind
of singularity is {\it not} realized in the Jacobs solutions (nor the Bianchi II with a massless scalar field, or ``stiff'' matter) \cite{collins}.
This can be understood easily. This singularity happens when $k_1>0$ and $k_2=k_3=0$ (cyclic on 1,2,3) in the Bianchi I case, satisfying the constraint equations:
$k_1+k_2+k_3=1$ that gives $k_1=1$. Now, the condition $k_1^2+k_2^2+k_3^2+k_\phi^2= 1$ implies that $k_\phi=0$, which means that there is no matter.  Thus, this type of singularity is not possible if there is stiff matter present.
\end{enumerate}

These names for the different types of singularities, introduced by Thorne in \cite{thorne}, refer to the change of the shape of a spherical element as the singularity is approached.
The Jacobs solutions will be useful in our analysis, since Bianchi II is past and future asymptotic to Bianchi I\footnote{This is true when the matter is a massless scalar field.
For other matter the asymptotic solutions are different.}
(see, for instance \cite{w_hsu,mac} and section 9.3 of \cite{wain_ellis}).
Additionally, Bianchi II is a limiting case for the effective equations of Bianchi II that come from its quantization \cite{bianchiII} in the
loop quantum cosmology (LQC) framework. 
Then, in the classical region, Bianchi I (Jacobs) solutions
are limiting cases for the effective Bianchi II that arises from LQC.\\

We will now rewrite the classical theory in terms of triads and connections in order to connect with the effective theory that 
comes from the quantum theory. To do this we use the fiducial triads and
co-triads and introduce a convenient parametrization of the phase
space variables, $E^a_i, A_a^i$ given by
\be  E^a_i = p_i L_i V_0^{-1} \sqrt{|\q|}\,\e^a_i \qquad \mathrm{and}
\qquad A_a^i = c_i L_i^{-1}\,\w_a^i, \ee
without sum over $i$, where $V_0=L_1L_2L_3$ is the fiducial volume and $L_i$ the fiducial lengths with respect to 
the fiducial metric $\q_{ab}:=\delta_{ij}\w_a^i \w_b^j$ with co-triads
\be
\w_a^1 =  (\d x)_a -\alpha z(\d y)_a, \quad \w_a^2 = (\d y)_a \quad \w_a^3 = (\d z)_a
\ee
and triads
\be
\e_1^a = \left(\f{\p }{\p x}\right)^a , \quad 
\e_2^a = \alpha z\left(\f{\p }{\p x}\right)^a + \left(\f{\p }{\p y}\right)^a, \quad 
\e_3^a = \left(\f{\p }{\p z}\right)^a ,
\ee
with Lie bracket $[\e_2,\e_3]=-\e_1$,\footnote{We are using the notation from \cite{bianchiII} and chapter 11 of \cite{hawking}. 
Note that this is not the typical choice ($[\e_2,\e_3]=\e_1 $) for Bianchi II. This choice only implies
a change from one invariant set ($n_1=1$) of Bianchi II to the other one ($n_1=-1$), but the physical properties
are the same, given that the equations of motion have this discrete symmetry. 
For more details see \cite{wain_ellis,ellis_mac,hawking}.}, with $\alpha=1$. 
A point in the phase space is now coordinatized by eight real numbers
$(p_i,c_i, \phi, p_\phi)$, with $\phi$ the scalar field and $p_\phi$ its conjugate momentum. 
The Poisson brackets are given by
\be \label{brackets}
\{c_i,\, p_j\} \, = \, 8\pi G \gamma \,\delta_{ij} \,
\hspace{1cm} \{\phi,p_\phi\}=1\, ,
\ee
where $\gamma$ is the Barbero-Immirzi parameter. 
The Hamiltonian formulation will be complete with the Hamiltonian constraint, which for 
a lapse function $N=\sqrt{|p_1p_2p_3|}$ reads \cite{bianchiII},
\be \label{classical-H}
\mathcal{C}_H=
\f{1}{8\pi G\gamma^2}\Bigg[ p_1p_2c_1c_2 + p_2p_3c_2c_3 + p_1p_3c_1c_3 + \alpha \epsilon p_2p_3c_1
-\left.(1+\gamma^2)\left(\f{\alpha p_2p_3}{2p_1}\right)^2 \right] - \f{p_\phi^2}{2} =0 
\ee
where again $\alpha$ distinguish between Bianchi I and Bianchi II, 
$\epsilon=\pm 1$ depending on whether the frame $e^a_i$ is right or left
handed (in our solutions we assume $\epsilon = 1$, i.e. $p_i > 0$).  
The equations of motion are given by the Poisson brackets with the Hamiltonian 
constraint
\ba
\dot{p_1}&=&\gamma^{-1}(p_1p_2c_2+p_1p_3c_3+\alpha\epsilon p_2p_3),\label{p1dot}\\
\dot{p_2}&=&\gamma^{-1}(p_2p_1c_1+p_2p_3c_3), \label{p2dot}\\
\dot{p_3}&=&\gamma^{-1}(p_3p_1c_1+p_3p_2c_2), \label{p3dot}\\
\dot{c_1}&=&-\gamma^{-1}\Big[p_2c_1c_2+p_3c_1c_3+\f{1}{2p_1}(1+\gamma^2)
\big(\f{\alpha p_2p_3}{p_1}\big)^2\Big], \\
\dot{c_2}&=&-\gamma^{-1}\Big[p_1c_2c_1+p_3c_2c_3+\alpha\epsilon
p_3c_1-\f{1}{2p_2} (1+\gamma^2)\big(\f{\alpha
p_2p_3}{p_1}\big)^2\Big], \\
\dot{c_3}&=&-\gamma^{-1}\Big[p_1c_3c_1+p_2c_3c_2+\alpha\epsilon p_2
c_1-\f{1}{2p_3}(1+\gamma^2)\big(\f{\alpha p_2p_3}{p_1}\big)^2\Big],\\
\dot{p_\phi}& = &0 \hspace{1cm}\Rightarrow p_\phi = {\rm constant} ,\\
\dot{\phi}& = &-p_\phi \hspace{0.5cm}\Rightarrow \phi = -p_\phi \tau ,
\ea
where $\tau$ is called the harmonic time
 (with lapse $N=\sqrt{|p_1p_2p_3|}$). The last equation shows that the field $\phi$ can be used
as internal time. The harmonic time $\tau$ is related to the cosmic time $t$ (with lapse $N=1$) by the equation
\be  \label{times_rel} 
\f{\d}{\d t}= \f{1}{\sqrt{|p_1p_2p_3|}} \, \, \f{\d}{\d\tau}\, .
\ee
It is with respect to this time that we will define the observable quantities. The derivative respect to the cosmic time will be denoted
by ${\d O}/{\d t}=O'$, then $O'=\dot O/\sqrt{|p_1p_2p_3|}$.
From the equations of motion it is straightforward to show that the classical solutions posses the constants of motion
\ba
c_1p_1+c_2p_2 &=:& \alpha_{12} \, ,\\
c_1p_1+c_3p_3 &=:& \alpha_{13} \, ,\\
c_3p_3-c_2p_2 &=& \alpha_{32} = \alpha_{13}-\alpha_{12} \label{alpha32}\, ,
\ea
with $\alpha_{12}, \alpha_{13}$ constants. This will allow us to solve analytically equations \eqref{p2dot} and \eqref{p3dot},
\ba
\dot p_2 = \gamma^{-1}p_2(p_1c_1 + p_3c_3)= \gamma^{-1}p_{2} \alpha_{13}  \qquad & \Rightarrow &\qquad
p_{2} = p_{2}^0\, \mbox{exp}\left(\f{\alpha_{13} \tau}{\gamma}\right) ,\\
\dot p_3 = \gamma^{-1}p_3(p_1c_1 + p_2c_2) =\gamma^{-1}p_{3} \alpha_{12}  \qquad & \Rightarrow &\qquad
p_{3} = p_{3}^0\, \mbox{exp}\left(\f{\alpha_{12} \tau}{\gamma}\right) .
\ea
The existence of this exact solutions gives us the opportunity to compare the numerical solutions with the analytical ones for $p_2$ and $p_3$. 
Also, we can check that during the evolution, for the numerical solutions, $\alpha_{12}$ and $\alpha_{13} $
remain  constant.\\

In order to determine how the classical singularities are resolved and how the effective equations evolve,
the quantities that we will study are:

\begin{enumerate}
\item The {\it directional scale factors} $$a_i=L_i^{-1}\sqrt{\f{p_jp_k}{p_i}},$$ with {\small $i \neq j\neq k\neq i $} and 
$p_i>0$, $i=1,2,3$. This makes easy the comparison between the classical and the effective solutions.

\item The {\it directional Hubble parameters }
$$H_i = \f{a_i'}{a_i} =
\f{1}{2}\left(\f{p_j'}{p_j}+\f{p_k'}{p_k}-\f{p_i'}{p_i}\right),$$
with {\small $i \neq j\neq k\neq i $}. This quantity tells us when each direction bounces ($H_i=0$) or if these 
directions are contracting ($H_i<0$) or expanding ($H_i>0$). 

\item The {\it expansion} $$\theta=\f{V'}{V}=H_1+H_2+H_3.$$ 

This quantity gives the total expansion rate and determines when there is a global bounce ($\theta=0$). An equivalent
quantity is the {\it mean Hubble parameter} $$H=\f{\theta}{3}=\f{1}{3}(H_1+H_2+H_3).$$ 

\item The {\it matter density} $$\rho=\f{p_\phi^2}{2V^2}=\f{p_\phi^2}{2p_1p_2p_3}.$$ If the singularities
are resolved then this quantity must be finite thoughout the evolution. Other quantity that measures the dynamical
importance of the matter content is the {\it density parameter} $\Omega$, defined by
\be
\Omega :=\f{ 8\pi G}{3}\f{\rho}{H^2}\, .
\ee
This parameter is related to the Kasner exponents in Bianchi I by the equation $\Omega=\f{3}{2}k_\phi^2 = \f{3}{2}(1-k_1^2-k_2^2-k_3^2)$  (see, for instance \cite{jacobs}).

\item The {\it shear} 
$$\sigma^2=\sigma_{ab}\sigma^{ab}=\f{1}{3}[(H_1-H_2)^2+(H_1-H_3)^2+(H_2-H_3)^2] = \sum_{i=1}^3 H_i^2 -\f{1}{3}\theta^2\, .$$ 
Note that this definition of $\sigma^2$ differs from the standard definition $\sigma^2=\f{1}{2}\sigma_{ab}\sigma^{ab}$.
Another important quantity is the {\it shear parameter}
\be
\Sigma^2 := \f{3\sigma^2}{2\theta^2}=\f{\sigma^2}{6H^2} ,
\ee

that measures the rate of shear (i.e. anisotropy) in terms of the expansion.
In Bianchi I the expansion satisfies $\theta =t^{-1}$ (using the fact that $V=a_1a_2a_3=t^{k_1+k_2+k_3}=t$, without putting explicitly the units). 
Then, the shear parameter reduces to the relation 
$\Sigma^2=\f{3}{2}t^2 \sigma^2=\f{3}{2}V^2\sigma^2=\f{3}{2}a^6\sigma^2=9 \Sigma^2_{\rm BI}$, 
where $\Sigma^2_{\rm BI}:=\f{1}{6}\sigma^2a^6$ was the shear parameter used in Bianchi I \cite{bianchiold, bianchiI}, with 
$a:=(a_1a_2a_3)^{1/3}$ the mean scale factor.

\item The {\it Ricci scalar} for the Bianchi II metric, Eq. (\ref{metric}), is given by
\be
R = 2\left(\f{a_1''}{a_1}+\f{a_2''}{a_2}+\f{a_3''}{a_3}+\f{a_1'a_2'}{a_1a_2}+\f{a_1'a_3'}{a_1a_3}+\f{a_2'a_3'}{a_2a_3}\right) - \alpha^2\f{a_1^2}{2a_2^2a_3^2} \, .
\ee

When $\alpha=0$ it reduces to the Ricci scalar for Bianchi I. In terms of the new variables $(c_i,p_i)$ it has a simple expression
\be
R=\f{{p}_1''}{p_1}+\f{{p}_2''}{p_2}+\f{{p}_3''}{p_3}-\f{1}{2}x^2\, ,
\ee
where $x=\alpha\sqrt{\f{p_2p_3}{p_1^3}}$. This equation can be rewritten in terms of other observables,
\be\label{Ricci}
R = 2\theta'+\sigma^2 + \f{4}{3}\theta^2 -\f{1}{2}x^2 \,.
\ee
This relation provides the easiest way to calculate the Ricci scalar numerically. 
On the classical solutions $\theta'$ is given by the Raychaudhuri equation
\be
\theta' = -\f{1}{2}\theta^2-\sigma^2-16\pi G \rho\, .
\ee

One important feature of the Bianchi II model is that the spatial curvature is different from zero. thu,  we can introduce other quantity
that give us information about the dynamical contribution due to the extrinsic curvature, namely the {\it curvature parameter} $K$, given by
\be
K = \f{3 x^2}{4\theta^2}= \f{x^2}{12 H^2}.
\ee
Our choice of $\Omega, \Sigma^2$ and $K$ is motivated by the fact that, on the classical solutions for Bianchi II, they satisfy the equation
\be
\Omega + \Sigma^2 + K =1 \, .
\ee
These parameters have the `problem' that they are infinity at the bounce ($\theta=0$) by definition, so they
are not very useful to explore that regime. But, since we are interested in its asymptotic behavior for large
volume, and the information they can give us there, this pathological behavior at the bounce is not relevant and does not reflect any problem with the singularity resolution.

\item The {\it Kasner exponents} 
\be \label{kasner}
k_i=\f{H_i}{|\theta|}.
\ee
These parameters are very useful to determine when the solutions have a Bianchi I behavior. 
We have taken the absolute value in $\theta$ because we want that different signs in $H_i$ 
specify if the directions are expanding ($k_i>0$) or contracting ($k_i<0$).
In order to prove Eq. (\ref{kasner}) we use that, in Bianchi I,  $a_i=t^{k_i}L_i$ (using 
explicitly the fiducial lenghts), then
\be
k_i = \f{a_i' t^{1-k_i}}{L_i} = \f{a_i' t}{t^{k_i} L_i} = \f{a_i' t}{a_i} = \f{a_i' V_0 t}{a_i V_0} 
= \f{a_i' V}{a_i V_0} = \f{a_i' V}{a_i V'} = \f{H_i}{\theta} \, , 
\ee
with $V=a_1a_2a_3=V_0t$.

\end{enumerate}

One important remark is that all
the previous expressions to calculate the observable quantities apply also to the effective solutions. The only difference is the calculation of
$\theta'$ in the effective theory, for which it is necessary to use the effective equations of motion.  
That will be shown in the next section. 

To complete the classical picture we give the relations between phase space variables and metric variables, 
which are given by
\be
p_1 = a_2a_3L_2L_3\, , \qquad
p_2 = a_1a_3L_1L_3\, , \qquad
p_3 = a_1a_2L_1L_2\, , \label{p_a}
\ee
and
\ba 
c_1 &=& \gamma L_1a_1H_1 +\f{\alpha}{2}\f{a_1^2L_1^2}{a_2a_3L_2L_3}\, , \label{c_a1} \\
c_2 &=& \gamma L_2a_2H_2 -\f{\alpha}{2}\f{a_1L_1}{a_3L_3}\, , \label{c_a2} \\
c_3 &=& \gamma L_3a_3H_3 -\f{\alpha}{2}\f{a_1L_1}{a_2L_2}\, , \label{c_a3}
\ea
assuming $a_i>0, p_i > 0$ and $\epsilon=1$. Relations (\ref{p_a}) are satisfied at the kinematical level whereas 
Eqs. (\ref{c_a1},\ref{c_a2},\ref{c_a3}) are satisfied at the dynamical level, i.e., on the space of solutions. These relations can be shown using the Hubble parameters  (with explicit fiducial lenghts),
\ba
H_i=\f{1}{a_i}\f{d a_i}{d t} &=&
\f{1}{2\sqrt{p_1p_2p_3}}\left(\f{\dot p_j}{p_j}+\f{\dot p_k}{p_k}-\f{\dot p_i}{p_i}\right)\\ 
\f{d a_i}{d t} &=&
\f{1}{2p_iL_i} \left(\f{\dot p_j}{p_j}+\f{\dot p_k}{p_k}-\f{\dot p_i}{p_i}\right) \, ,
\ea
with $i\neq j\neq k \neq i$. If we put the equations of motion (\ref{p1dot}, \ref{p2dot}, \ref{p3dot}) into this expression, we get
\ba
\f{d a_i}{d t} &=&
\f{c_i}{\gamma L_i} + {\rm fsign}(i) \f{1}{2p_i L_i}\f{\alpha}{\gamma}\f{p_2p_3}{p_1} \, ,
\ea
where fsign$(i)=-1$ if $i=1$ and fsign$(i)=1$ if $i=2,3$. Using the relations between $a_i$ and $p_i$, Eq. (\ref{p_a}), 
and the definition of the Hubble parameters we get
\ba
c_i &= &\gamma L_i \f{d a_i}{d t} - {\rm fsign}(i) \f{1}{2p_i}\alpha L_1^2a_1^2\\
 &= &\gamma L_i a_iH_i - {\rm fsign}(i)\f{\alpha}{2} \f{ L_1^2a_1^2}{L_jL_ka_ja_k} \, .
\ea
These are precisely Eqs. (\ref{c_a1}, \ref{c_a2}, \ref{c_a3}).


\section{Effective Dynamics}
\label{sec:3}

The effective theory was derived from the loop quantization of Bianchi II defined in reference \cite{bianchiII}, 
using the procedure outlined in \cite{vt}.
Taking a right-hand frame $e_i^a$ (i.e. $\epsilon = 1$) and the lapse function $N=\sqrt{p_1p_2p_3}$, 
the effective Hamiltonian constraint is given by,
\begin{align} 
\mathcal{C}_H & =
\f{p_1p_2p_3}{8\pi G\gamma^2\lambda^2}
\left[\f{}{}\sin\bar\mu_1c_1\sin\bar\mu_2c_2+\sin\bar\mu_2c_2
\sin\bar\mu_3c_3+\sin\bar\mu_3c_3\sin\bar\mu_1c_1\right] \nonumber\\ 
& \quad + \f{1}{8\pi G\gamma^2}
\Bigg[\f{\alpha(p_2p_3)^{3/2}}{\lambda\sqrt{p_1}}\sin\bar\mu_1c_1 
-(1+\gamma^2)\left(\f{\alpha p_2p_3}{2p_1}\right)^2 \Bigg] - \f{p_\phi^2}{2} =0 \, ,  \label{effective-H}
\end{align}
with 
\be
\bar\mu_1 = \lambda\sqrt{\f{p_1}{p_2 p_3}} , \qquad
\bar\mu_2 = \lambda\sqrt{\f{p_2}{p_1 p_3}} , \qquad
\bar\mu_3 = \lambda\sqrt{\f{p_3}{p_1 p_2}} .
\ee
The value of $\lambda$ is chosen such that $\lambda^2=\Delta$ corresponds to the minimum eigenvalue of the
area operator in loop quantum gravity (corresponding to an edge of ``spin $1/2$''). With this choice the free parameter becomes $\Delta=4\sqrt{3}\pi\gamma\lp^2$. 
Since $\sin\bar\mu_ic_i\le 1$ the matter density $\rho=\f{p_\phi^2}{2V^2}=\f{p_\phi^2}{2p_1p_2p_3} $ satisfies
\be 
\rho\le \f{3}{8\pi G\gamma^2\lambda^2}
+\f{1}{8\pi G \gamma^2 } \left[ \f{x}{\lambda}-\f{(1+\gamma^2)x^2}{4}\right]\, ,
\quad {\rm with}\quad x=\alpha\sqrt{\f{p_2p_3}{p_1^3}}. 
\ee
The maximum of the expression in square brackets is attained at 
$x=\f{2}{(1+\gamma^2)\lambda}\approx 0.83$, then
\be 
\rho_{\rm matt} \lesssim 1.315 \rho_{\rm crit} \approx 0.54 \rho_{\rm Pl}\, ,
\ee
with $\rho_{\rm crit}=\f{3}{8\pi G\gamma^2\lambda^2}\approx 0.41 \rho_{\rm Pl}$ 
the critical density found in the isotropic case \cite{aps} and $\rho_{\rm Pl}=m_{\rm Pl}/\lp^3$ is the Planck density.
This shows that the matter density is bounded.
This bound is higher than the one found in  Bianchi I $ \rho_{\rm matt}\lesssim 0.41 \rho_{\rm Pl}$ 
and the isotropic case  $\rho_{\rm matt}\approx 0.41 \rho_{\rm Pl}$. 
Furthermore, the density in all the solutions in Bianchi I with shear different from zero has a bounce density less 
than its value in the isotropic solution. Then there is an open question: Are there generic solutions in which 
the matter density is larger that its value in the isotropic solutions? We will show that the answer is in the affirmative, which leaves us with another open question: How do we find the solutions that saturate the matter density? 
These kind of solutions are shown in section \ref{sec:4:LRS}.

If we set $\alpha=0$ into  Eq. \eqref{effective-H} we recover the Hamiltonian constraint for Bianchi I \cite{bianchiI}. Also,  if we take the Bianchi II case, $\alpha = 1$, we can recover Bianchi I as a limiting case when $x\rightarrow 0$, 
or equivalently $ p_1^3 \gg p_2p_3$ (in metric variables this condition is expressed as $a_2a_3 \gg a_1$). 
It is important to have in mind that the Bianchi I model is a limiting case and is not contained within the Bianchi II model. 

The equations of motion for the effective theory are given by Poisson brackets with the Hamiltonian constraint,
\ba 
\dot{p_1} &=& \f{p_1^2}{\gamma\bar\mu_1}\left(\sin\bar\mu_2c_2+
\sin\bar\mu_3c_3+\lambda x \right)\cos\bar\mu_1c_1, \\
 \dot{p_2} &=& \f{p_2^2}{\gamma\bar\mu_2}(\sin\bar\mu_1c_1+\sin\bar\mu_3c_3)
\cos\bar\mu_2c_2, \\
 \dot{p_3} &=& \f{p_3^2}{\gamma\bar\mu_3}(\sin\bar\mu_1c_1+\sin\bar\mu_2c_2)
\cos\bar\mu_3c_3, 
\ea
\begin{align} 
\dot{c_1} &= -\f{p_2p_3}{2\gamma\lambda^2}\left[\f{}{}
2(\sin\bar\mu_1c_1\sin\bar\mu_2c_2+\sin\bar\mu_1c_1\sin\bar\mu_3c_3
+\sin\bar\mu_2c_2\sin\bar\mu_3c_3) \nonumber \right.\\ 
& \qquad +{\bar\mu_1c_1}\cos\bar\mu_1c_1
(\sin\bar\mu_2c_2+\sin\bar\mu_3c_3)-{\bar\mu_2c_2}\cos\bar\mu_2c_2
(\sin\bar\mu_1c_1+\sin\bar\mu_3c_3) \nonumber \\ 
& \qquad -{\bar\mu_3c_3}\cos\bar\mu_3c_3
(\sin\bar\mu_1c_1+\sin\bar\mu_2c_2)+{\lambda^2x^2}(1+\gamma^2) \nonumber \\ 
& \left. \qquad +{\lambda x}(\bar\mu_1c_1\cos\bar\mu_1c_1-\sin\bar\mu_1c_1)\f{}{}\right], 
\end{align}
\begin{align} 
\dot{c_2} &= -\f{p_1p_3}{2\gamma\lambda^2}\left[\f{}{}
2(\sin\bar\mu_1c_1\sin\bar\mu_2c_2+\sin\bar\mu_1c_1\sin\bar\mu_3c_3
+\sin\bar\mu_2c_2\sin\bar\mu_3c_3) \nonumber \right.\\ 
& \qquad -{\bar\mu_1c_1}\cos\bar\mu_1c_1
(\sin\bar\mu_2c_2+\sin\bar\mu_3c_3)+{\bar\mu_2c_2}\cos\bar\mu_2c_2
(\sin\bar\mu_1c_1+\sin\bar\mu_3c_3) \nonumber \\ 
& \left.\qquad -{\bar\mu_3c_3}
\cos\bar\mu_3c_3(\sin\bar\mu_1c_1+\sin\bar\mu_2c_2)\right]
-{\lambda^2x^2}(1+\gamma^2) \nonumber \\ 
& \left.\qquad -{\lambda x}(\bar\mu_1c_1\cos\bar\mu_1c_1 -3\sin\bar\mu_1c_1) \f{}{}\right], 
\end{align}
\begin{align} 
\dot{c_3} &= -\f{p_1p_2}{2\gamma\lambda^2}\left[\f{}{}
2( \sin\bar\mu_1c_1\sin\bar\mu_2c_2+\sin\bar\mu_1c_1\sin\bar\mu_3c_3
+\sin\bar\mu_2c_2\sin\bar\mu_3c_3) \nonumber \right.\\ 
& \qquad -{\bar\mu_1c_1}\cos\bar\mu_1c_1
(\sin\bar\mu_2c_2+\sin\bar\mu_3c_3)-{\bar\mu_2c_2}\cos\bar\mu_2c_2
(\sin\bar\mu_1c_1+\sin\bar\mu_3c_3) \nonumber \\ 
& \left.\qquad +{\bar\mu_3c_3}
\cos\bar\mu_3c_3(\sin\bar\mu_1c_1+\sin\bar\mu_2c_2)\right]
-{\lambda^2x^2}(1+\gamma^2) \nonumber \\ 
& \left.\qquad -{\lambda x}(\bar\mu_1c_1\cos\bar\mu_1c_1 -3\sin\bar\mu_1c_1) \f{}{}\right],
\end{align}
Finally, for the matter we have
\ba
\dot{p_\phi}& = &0 \qquad \Rightarrow\qquad p_\phi = {\rm constant} ,\\
\dot{\phi}& = &-p_\phi \qquad\Rightarrow\qquad \phi = -p_\phi \tau .
\ea
Note that the equations for the matter part are equal to the classical ones, so the field $\phi$ also
plays the role of internal time. The field momentum $p_\phi$ is conserved, and its value coincides with the 
classical value. From the triad equations ($\dot{p_i}$) we can see that the bounce in direction $i$ occurs 
when $\cos(\bar\mu_i c_i)=0$, which can be satisfied at different times for each direction. 
These assertions do not imply that there is more than one bounce
in the matter density; the density only bounces one time, which is when the expansion $\theta$ is zero. This is 
called the {\it global} bounce. \\

We can use the equations of motion to give an explicit formula for $\theta'$ (derivative of the expansion with respect to cosmic time),
which is necessary to calculate the Ricci scalar. It is straightforward to show that 
\ba
\theta' &= &\f{1}{2\gamma\lambda} \left\{
\sum_{i=1}^3 \left[ 2\sin\bar\mu_ic_i + \cos\bar\mu_ic_i(\sin'\bar\mu_jc_j + \sin'\bar\mu_kc_k) \right] \right.\nonumber \\
& & \quad \quad\left. + \lambda x + \f{\lambda x }{2}\cos\bar\mu_1c_1 
\left( \f{p_2'}{p_2}+\f{p_3'}{p_3}-\f{3p_1'}{p_1} \right)\right\}\,, \label{thetadot}
\ea
where $j\neq i \neq k \neq j$ and $\sin'(\bar\mu_ic_i)=\cos(\bar\mu_ic_i)[\bar\mu_i'c_i+\bar\mu_ic_i']$, with $\bar\mu_i'=-\bar\mu_iH_i$.
It would be interesting if one could rewrite Eq. \eqref{thetadot} in terms of observable quantities ($H_i, \theta, \rho, \sigma^2$),
which would represent a generalization of the Raychaudhuri equation. 
Also, from the equations for $\dot{p_i}$ and $\dot{c_i}$ we get that
\be
c_3p_3-c_2p_2 =: \alpha_{32} 
\ee
is conserved and its value is the same than the classical one, as given by Eq. \eqref{alpha32}. 
The conserved quantities  ($p_\phi$ and $\alpha_{32}$) can be used to check that numerical solutions are evolving correctly  on the constraint surface.


\section{Numerical solutions}
\label{sec:4}

In this section we show the numerical solutions for the Bianchi II model. These equations admit different limits than can 
be used to check the accuracy of the solutions and explore the new insights that Bianchi II offers. In order to systematically study 
the solutions we need to have in mind the following facts:

\begin{enumerate}
\item In all the solutions, the constraint (or equivalently $p_\phi$) and $\alpha_{32}=c_3p_3-c_2p_2$ are conserved quantities.
\item In the classical solutions $\alpha_{12}$ and $\alpha_{13}$ are conserved quantities. 
\item When $\alpha = 0$ or $x=\sqrt{p_2p_3/p_1^3}\rightarrow 0$ (with $\alpha=1$), Bianchi II reduces to Bianchi I.
\item When $p_1=p_2=p_3$ and  $c_1=c_2=c_3$, Bianchi I reduces to the isotropic case.
\item In the isotropic case all the solutions to the effective equations have a maximal density equal to the critical density 
$\rho_{\rm crit}=\f{3}{8\pi G\gamma^2\lambda^2}\approx 0.41 \rho_{\rm Pl}$.
\item In the Bianchi I limit we will expect a maximal density less than the critical density.
\item Classical and effective solutions must be equal far away from the bounce. 
In fact the effective solutions to Bianchi I must connect two classical solutions with Kasner exponents related by
\be
k_1,k_2,k_3 \rightarrow k_1-\f{2}{3},k_2-\f{2}{3},k_3-\f{2}{3}
\ee
as was shown by Choiu \cite{bianchiold}.
\item In the Bianchi II model we expect a maximal density less than $1.315 \rho_{\rm crit}$.
\item The classical solutions diverge.
\item The numerical solutions must converge.

\end{enumerate}

Using these facts we shall now explore the Bianchi II solutions. We start from the classical limit showing that 
the effective solutions have a bounce and reduce to the classical ones far away from the bounce. Next, we explore the isotropic limit included into
the Bianchi I limit when there are no anisotropies. Later on,  we shall add anisotropies to the Bianchi I limit and show   that they reproduce the known solutions \cite{bianchiold}. Then we pass to the Locally Rotationally Symmetric (LRS) model of Bianchi II  and explore how to find the solutions with maximal density at the bounce. Finally we study the solutions in the vacuum limit with maximal shear.

To perform the analysis, solutions are plotted as functions of the cosmic time. All the integrations were made using a Runge-Kutta 4 method\footnote{The program is available by request.}. 
The units are: 
$\hbar =1, c=1, G=1, \gamma = 0.23753295796592, L_1 = 1, L_2 = 1, L_3 =1, \epsilon = 1$, the time step used is $dt=5\times 10^{-6}$. 
In all figures the density is plotted with normalization $\rho /\rho_{\rm crit}$.


\subsection{Comparison Between Classical and Effective Solutions}
\label{sec:4:class}

\begin{figure}[tpb!]
\begin{center}
\begin{tabular}{ll}
\includegraphics[width=8cm]{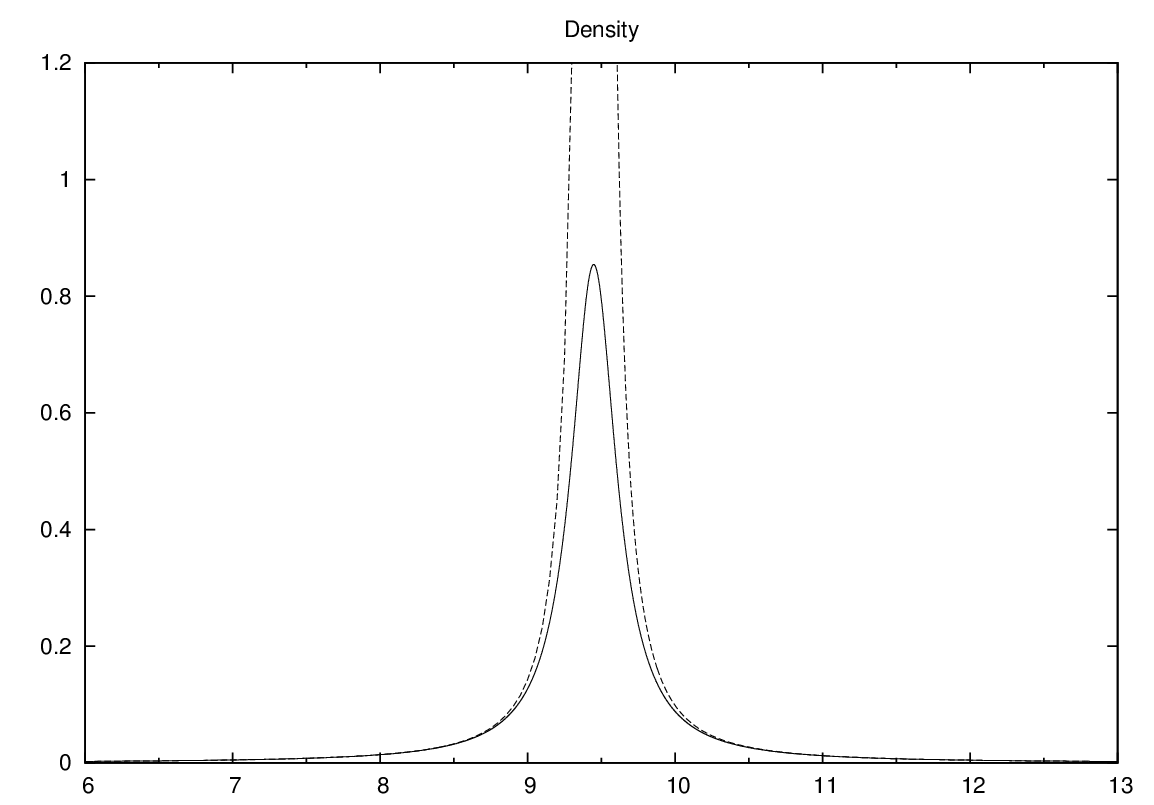}&
\includegraphics[width=8cm]{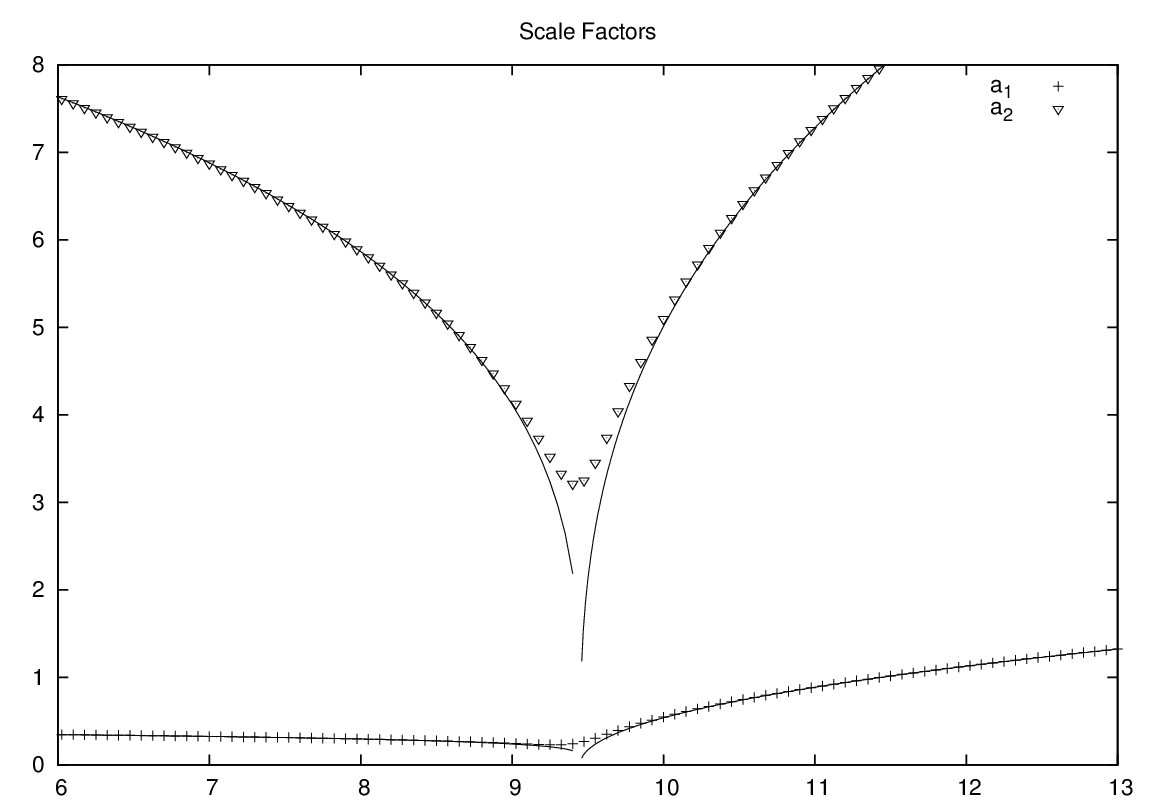}
\end{tabular}
\caption{Comparison between classical and effective solutions. Here it can be appreciated that near the big bang 
the classical solutions have an infinite density because the 
scale factors go to zero.}
\label{fig:class}
\end{center}
\end{figure}

In figure \ref{fig:class} the density and two scale factors 
(the third one is not shown for visualization purposes) are compared for the classical and effective solutions,
it is clear that, far from the bounce, the classical and effective solutions agree.  
Near to the bounce we can see that the classical density diverges in a finite time while 
the effective one bounces. This also happens to the shear (that is not shown in the plot). The finiteness of these quantities 
is due to the bounce in the scale factors that now are not going to zero 
in a finite time (there are solutions where some scale factors do not bounce and continue approaching zero,
but they need an infinite time to do so). This illustrates the manner in which the classical singularities are resolved. 

We understand by singularity resolution in the effective framework  the possibility to evolve the solutions for an arbitrary  time and that the solutions 
remain finite, thus signaling that the geodesics are inextendible. Classically, for homogeneous and anisotropic universes, the
singularities are present when the scale factor goes to zero in a finite time. This is related with an infinite density (expansion and shear) and  
incompleteness of geodesics. If the density, expansion and shear remain finite and can be evolved for any time then we say that singularities 
are resolved, i.e., the scale factors are not equal zero in a finite time or, equivalently, the geodesics are inextendible \cite{param,geometric}.

The initial conditions for effective and classical solutions at $t=100$ are: 
$c_1 = 0.01, c_2 = 0.02, c_3 = 0.03, p_1 = 10000, p_2 = 2000, p_3 = 200$ (it is evolved back in time), 
the field momentum $ p_\phi = 108.9$ is calculated from the Hamiltonian constraint. 
The initial conditions for the other classical solution at $t=0$ are: 
$c_1 = -1.920756\times 10^{-3}, c_2 = -7.971627\times 10^{-2}, c_3 = -19.61028, p_1 = 15864.53, p_2 = 635.5782, p_3 = 4.317425$ (and it is evolved forward in time).


\subsection{Isotropic Limit}

\begin{figure}[tbp!]
\begin{center}
\begin{tabular}{ll}
\includegraphics[width=8cm]{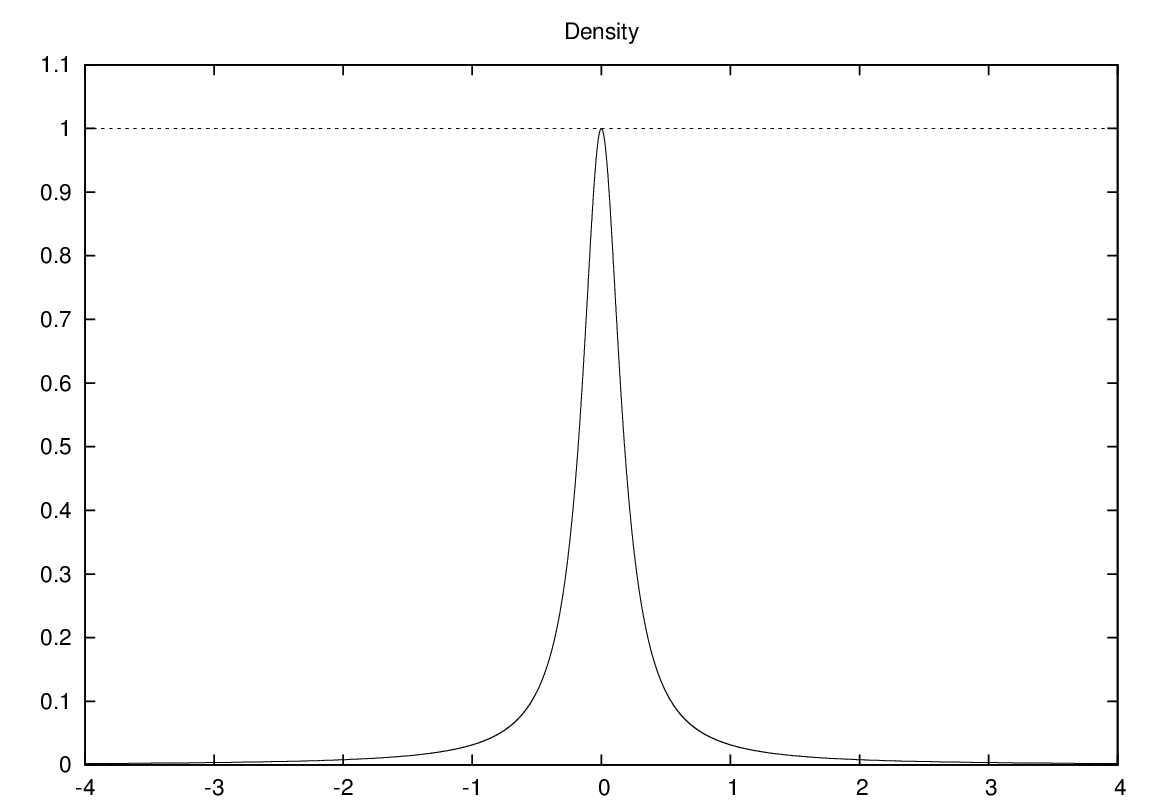}&
\includegraphics[width=8cm]{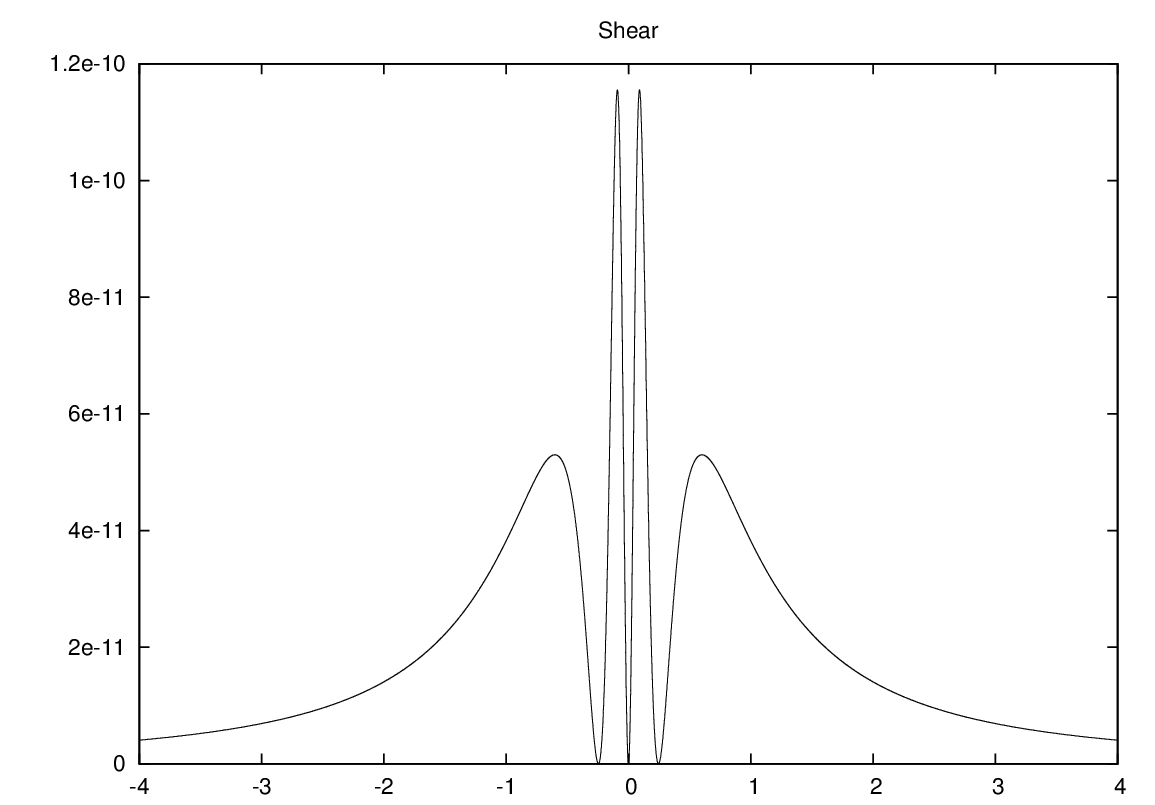}\\
\includegraphics[width=8cm]{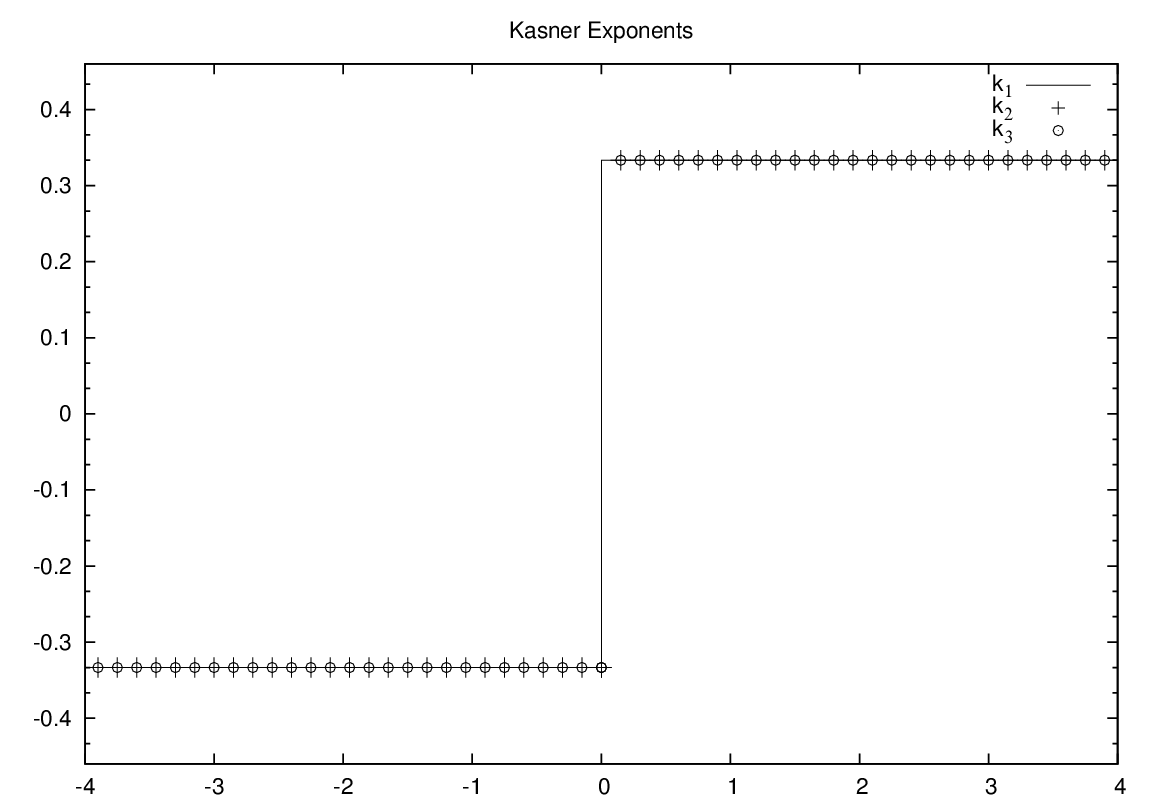}&
\includegraphics[width=8cm]{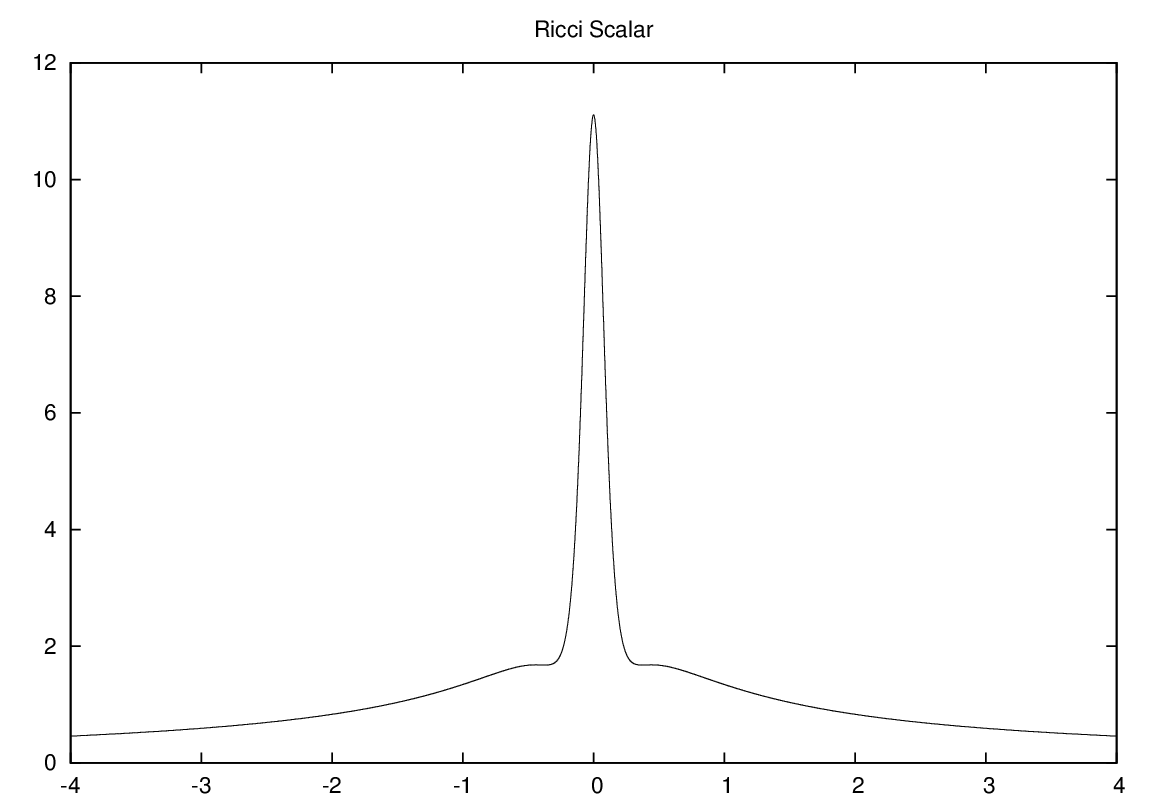}
\end{tabular}
\caption{Isotropic limit. The density at the bounce is $\rho /\rho_{\rm crit}=1$;
the shear is close to zero; the Kasner exponents are equal and change from $1/3$ to $1/3-2/3=-1/3$
when they are evolved back in time; the Ricci scalar strongly changes its behavior near to the bounce and 
have its maximal value at the bounce.}
\label{fig:iso}
\end{center}
\end{figure}

When $x\rightarrow 0$ Bianchi II has Bianchi I as a limit. Furthermore,
in this limit when $p_1=p_2=p_3$ and  $c_1=c_2=c_3$ Bianchi I reduced to the isotropic case. 
Bianchi II solutions near the isotropic limit are shown in figure \ref{fig:iso}, where 
we can see that solutions to the effective equations have a maximal density equal to the critical density $\rho_{\rm crit}$ 
\cite{aps2,slqc}. The shear is close to zero (in this case it is less than $1.2\times 10^{-10}$ in Planck units) 
and presents a non trivial behavior because it has four maxima and
vanishes when the density bounces. This is the first solution that is completely different from the known solutions in the isotropic and Bianchi I cases. 
The shear is small but not zero because this is not an isotropic solution, but it is only very close to it. One should keep in mind  that the  Bianchi I model and, therefore, 
the isotropic solutions are not contained within the Bianchi II solutions. \\

In order to have control on the isotropy, we set the initial conditions at the bounce, i.e. when 
$\bar\mu_1c_1=\bar\mu_2c_2=\bar\mu_3c_3= \pi/2$ (if the solution is isotropic the three directions bounce at same time, but the opposite is not true)
and then evolve back and forward in time, 
with these relations we only need three additional initial 
conditions, that can be either $(c_1, c_2, c_3)$ or $(p_1,p_2,p_3)$ or a combination of them. 
We take $p_1=1\times10^5,\, p_2=p_3=1\times10^3$ (which satisfy that $x\rightarrow 0$, i.e., $p_1^3 \gg p_2p_3$)
and using the fact that we are starting at the bounce ($\bar\mu_ic_i=\pi/2$) we calculate 
$c_1=2.185, c_2=c_3=2.185\times10^2$
and from the Hamiltonian constraint $p_\phi = 2.86\times10^5 $. 

Note that in this case it is not true that $p_1=p_2=p_3$ and  $c_1=c_2=c_3$. Then, one might why is this solution isotropic? 
The answer is that $c_1$ and $p_1$ are a rescaling of $c_{2,3}$ and $p_{2,3}$ that can be translated into a rescaling 
of the scalar factors such that the real criteria to say that it is isotropic are that the relations $a_1/a_2, a_2/a_3, a_1/a_3$ 
remain constant and $H_1=H_2=H_3$, which are satisfied by our initial conditions. 
This can be imagined like an isotropic universe described by the evolution of a fiducial cuboid.
The best way to see that this is an isotropic solution is to look at the Kasner exponents (Fig. \ref{fig:iso}), which 
are the same $k_1=k_2=k_3=\pm 1/3$. The different signs 
specify when the directions are expanding ($k_i>0$) or contracting ($k_i<0$).


\subsection{Bianchi I Limit}

\begin{figure}[tbp!]
\begin{center}
\begin{tabular}{ll}
\includegraphics[width=8cm]{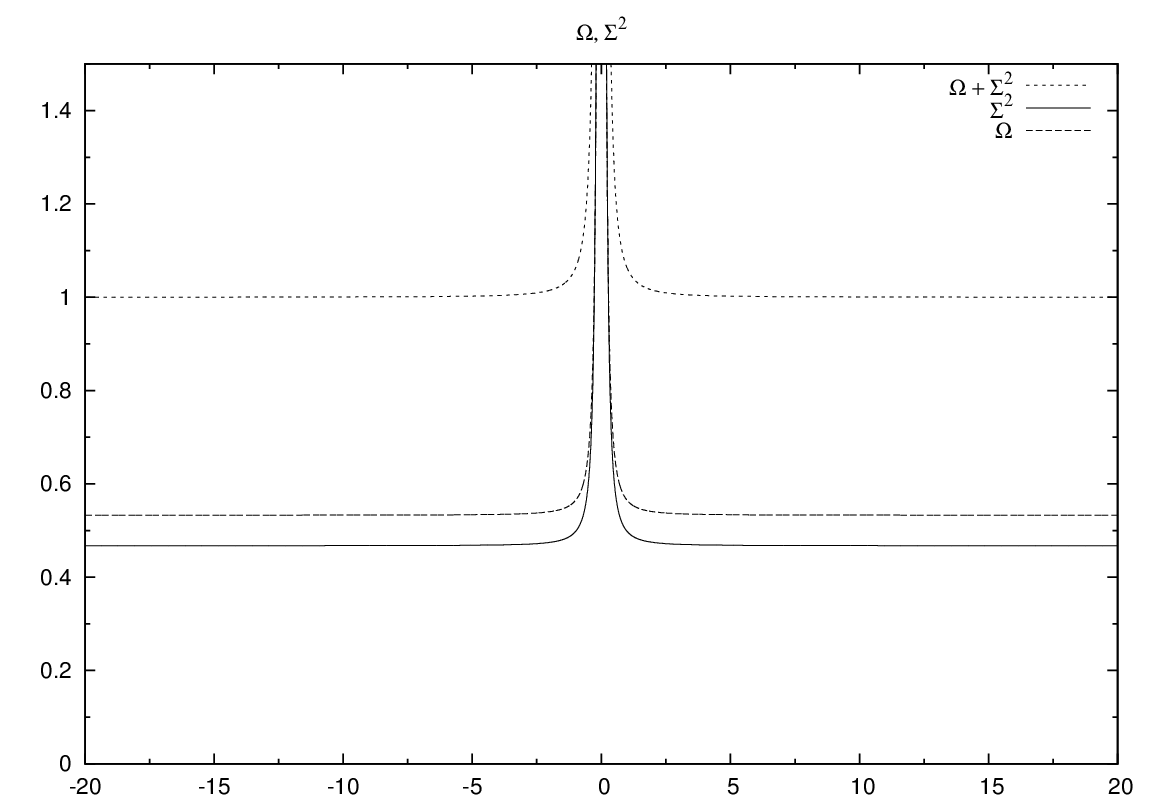}&
\includegraphics[width=8cm]{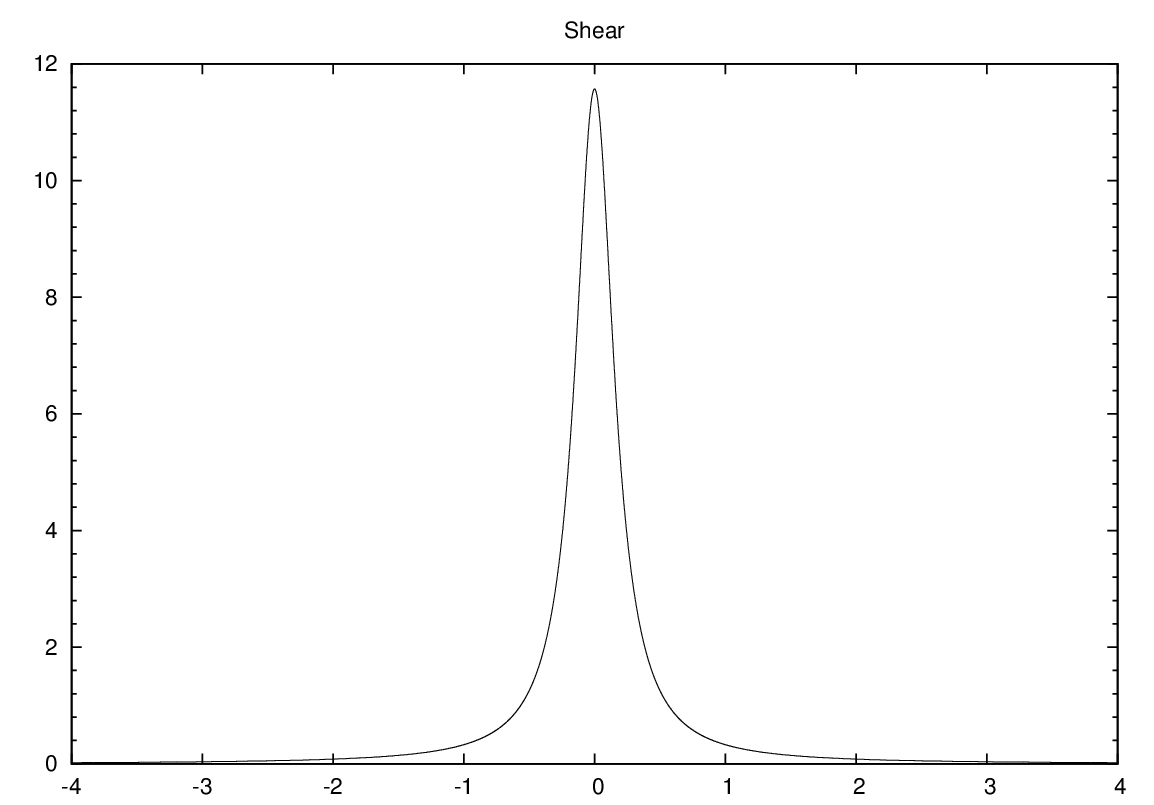}\\
\includegraphics[width=8cm]{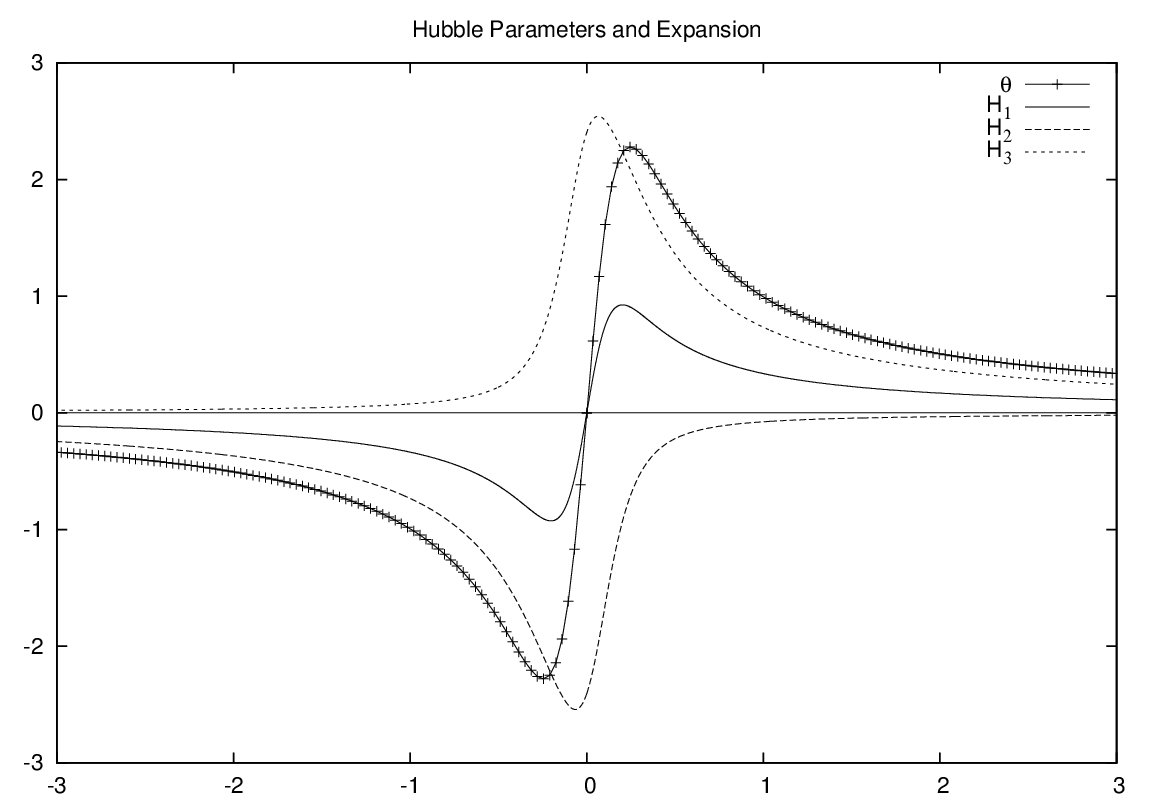}&
\includegraphics[width=8cm]{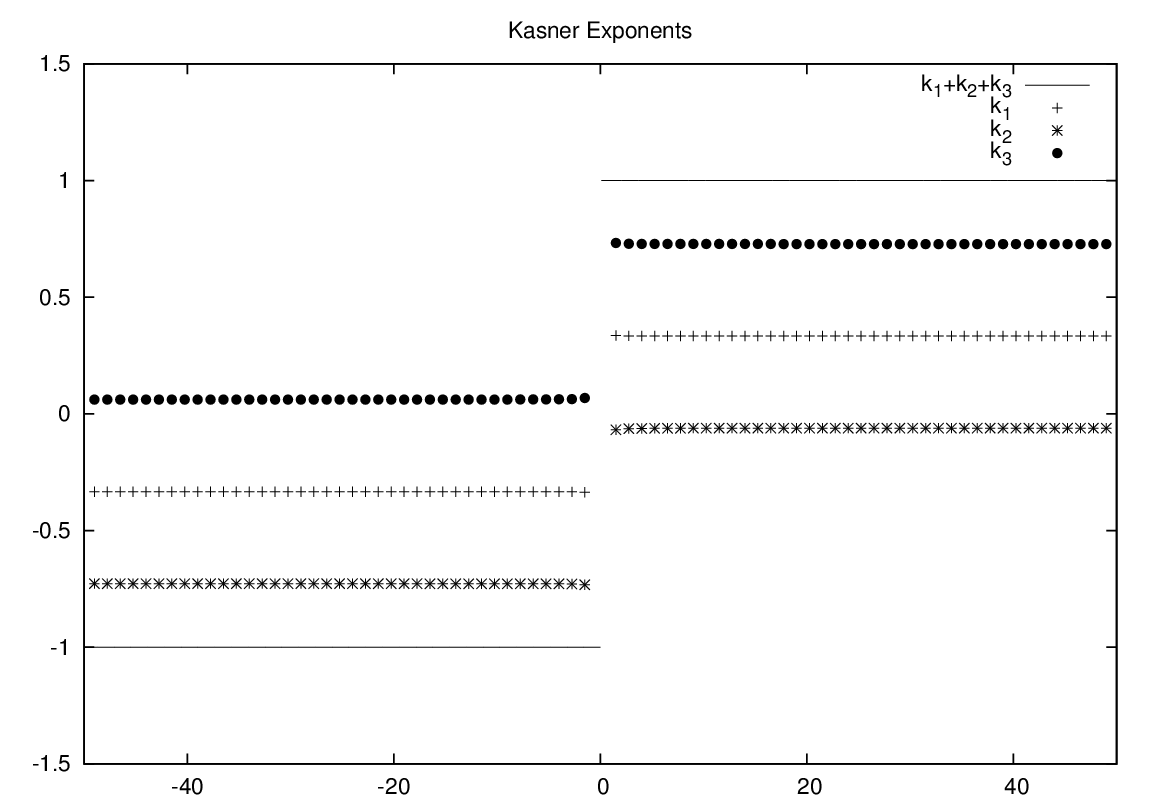}\\
\includegraphics[width=8cm]{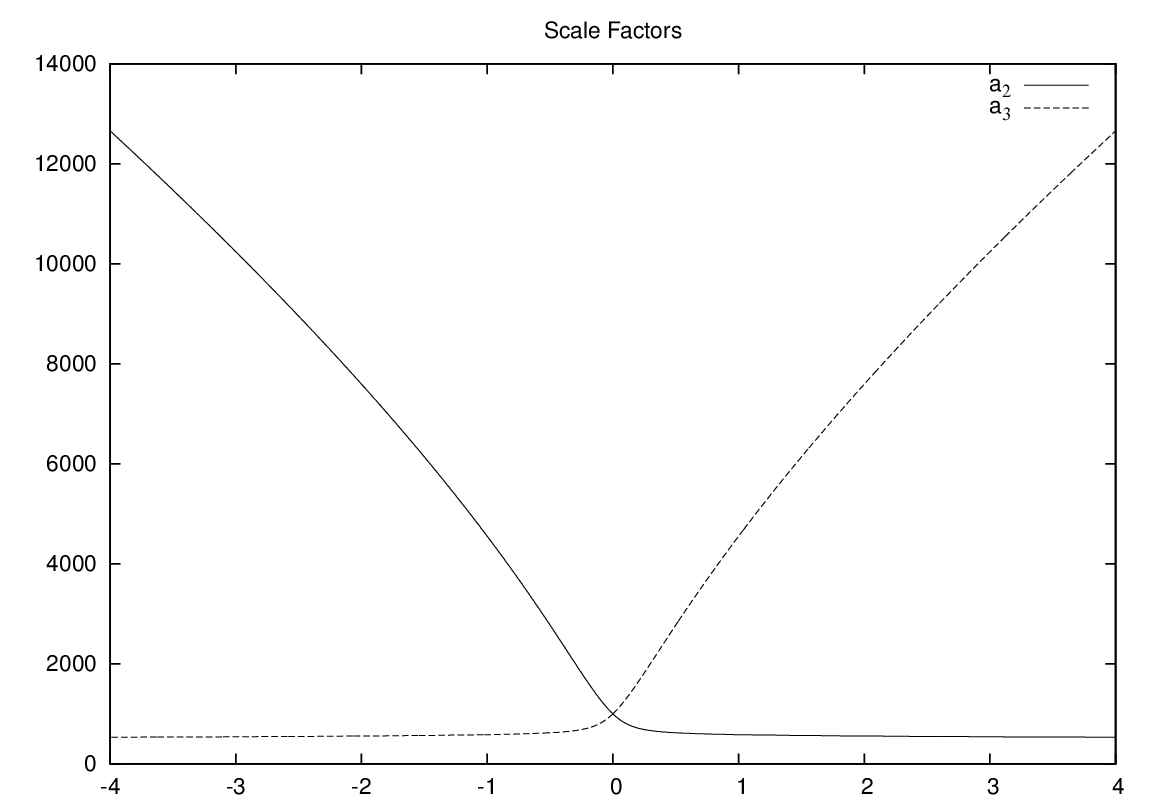}&
\includegraphics[width=8cm]{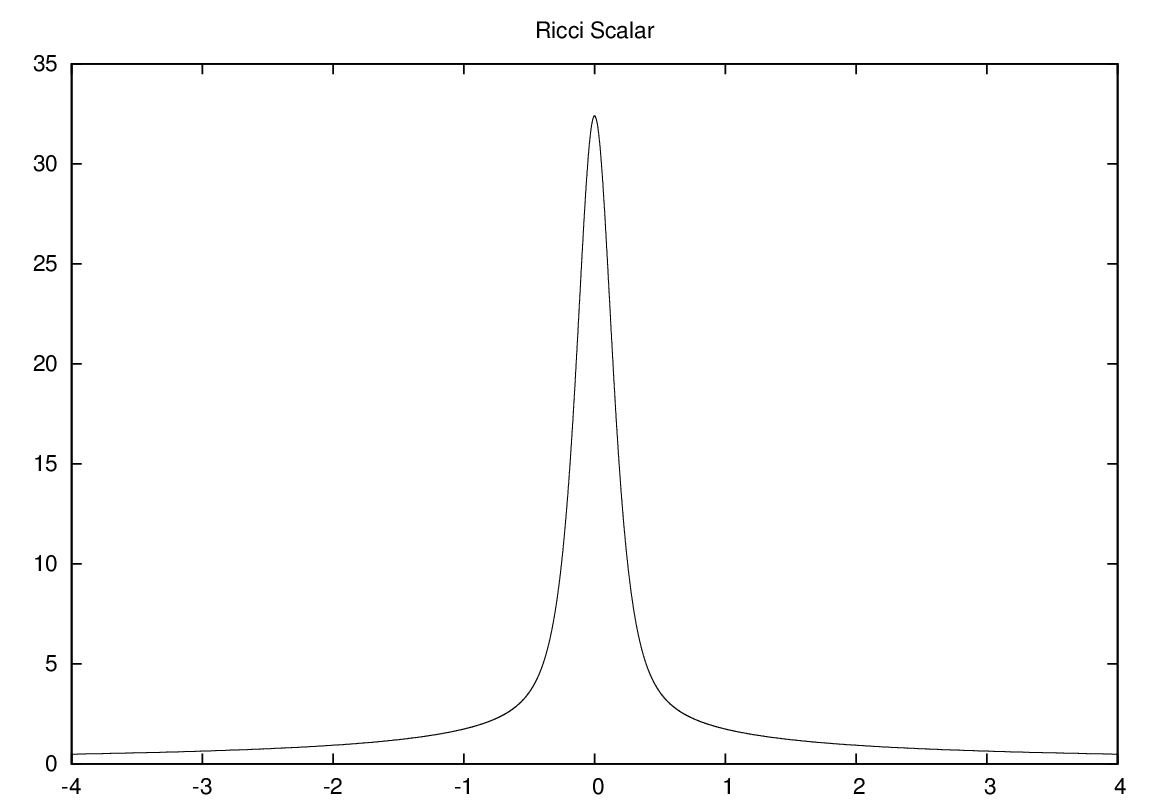}
\end{tabular}
\caption{Bianchi I limit. The density and shear have a dynamical contribution ($\Omega\neq 0, \Sigma^2\neq 0$);
the shear reaches its maximal value (for bianchi I) at the bounce; 
the Kasner exponents change from $k_i$ to $k_i-2/3$ when they are evolved back in time, as was shown by \cite{bianchiold};
one direction is expanding ($a_3$), one contracting ($a_2$) and one bounces ($a_1$);
the expansion has one zero (bounce) and it is finite all the time; the Ricci scalar presents a slower changes in its behavior in comparison to the isotropic
limit and it is finite too.}
\label{fig:BI}
\end{center}
\end{figure}

When $x\rightarrow 0$ the Bianchi II model has Bianchi I as a limit, which has been extensively studied \cite{bianchiold} and can be used as a reference point. 
In this limit we expect a maximal density less than the critical density  $\rho_{\rm crit}$ and a non zero shear with a maximal value
at the bounce, also $\Sigma^2$ and $\Omega$ must be conserved in the classical region (where $\Sigma^2+\Omega=1$, since $K=0$ in Bianchi I).
The Kasner exponents must satisfy the constraint equations for Bianchi I
solutions ($k_1+k_2+k_3=\pm 1$ and $k_1^2+k_2^2+k_3^2+k_\phi^2= 1$) and must change like 
$k_i \rightarrow k_i-2/3$ when  evolved back in time, as was shown in \cite{bianchiold}.
All these facts are shown in figure \ref{fig:BI}, which tells us that in the Bianchi I limit, the effective dynamics of Bianchi II reproduces the know behavior for Bianchi I.

A new and important feature of this solution is that it presents a maximal value of the shear for Bianchi I,
as can be seen in figure \ref{fig:BI}, where the value of shear at the bounce is $11.57=\f{10.125}{3\gamma^2\lambda^2}$, as reported in \cite{singh}. Moreover, 
it can be noticed that just one direction $a_1$ bounces and the other two directions $a_2,a_3$ do not, 
i.e., $H_2,H_3$ are not zero in a finite time. This implies
that these directions continue going to zero (or infinite) but now they need an infinite time to reach these values.
Note that in the classical region not all the directions are expanding (or contracting).
 This kind of universes have a classical singularity too, but it is a cigar-like singularity and it is different from 
the one in which all the directions are contracting (or expanding), called point-like singularity. Our simulations then show that this kind of singularity
is resolved too, with the notion of `singularity resolution' as explained in section \ref{sec:4:class}.

In this case the choice of initial conditions is: $\bar\mu_1c_1=\pi/2$, $\bar\mu_2c_2= \pi/6$, $\bar\mu_3c_3= 5\pi/6$ and $p_1=1\times 10^6$, $p_2=100$, $p_3=100$
(which imply that $x=\sqrt{p_2p_3/p_1^3}\approx 0$). With these initial conditions we get 
$c_1=0.069$, $c_2=230.28$, $c_3=  1151.39$ and $p_\phi=5.84\times10^4$. The initial time is at the global bounce ($\theta=0$), from which
 the solution is evolved back and forward in time.\\

\subsection{Locally Rotationally Symmetric (LRS) Solutions to Bianchi II}
\label{sec:4:LRS}

\begin{figure}[tbp!]
\begin{center}
\begin{tabular}{ll}
\includegraphics[width=8cm]{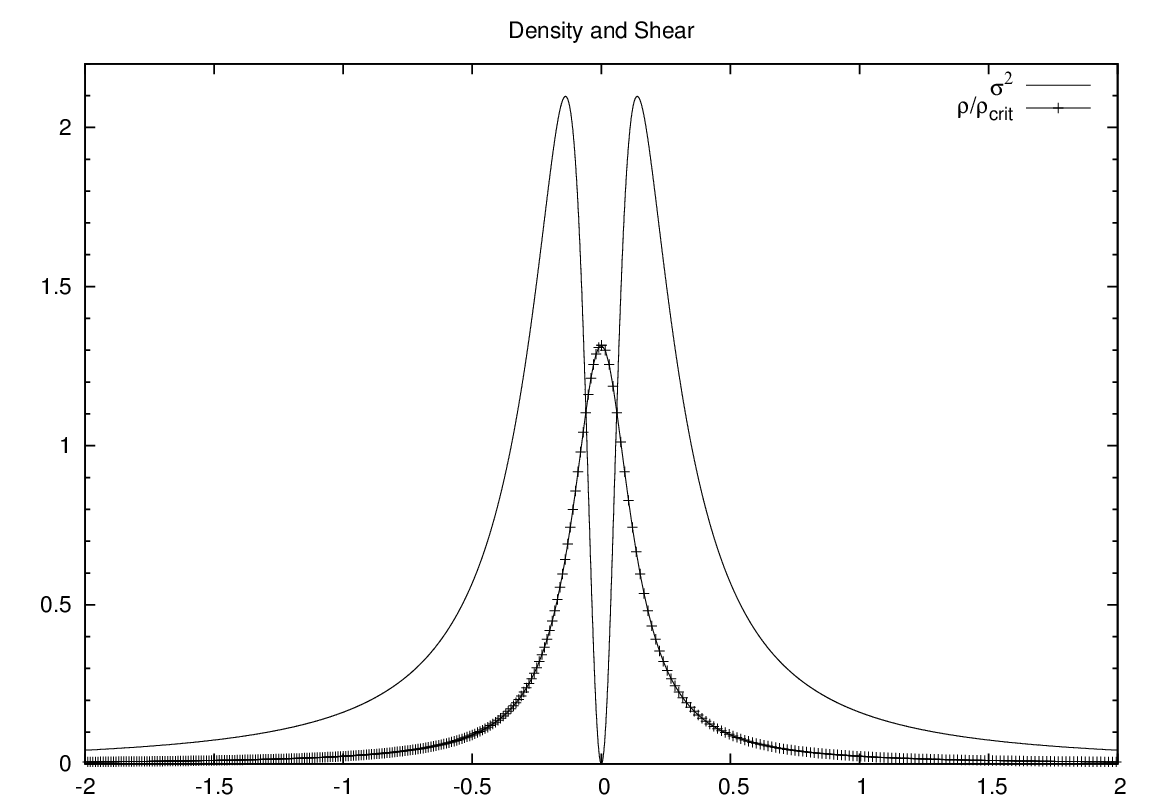}&
\includegraphics[width=8cm]{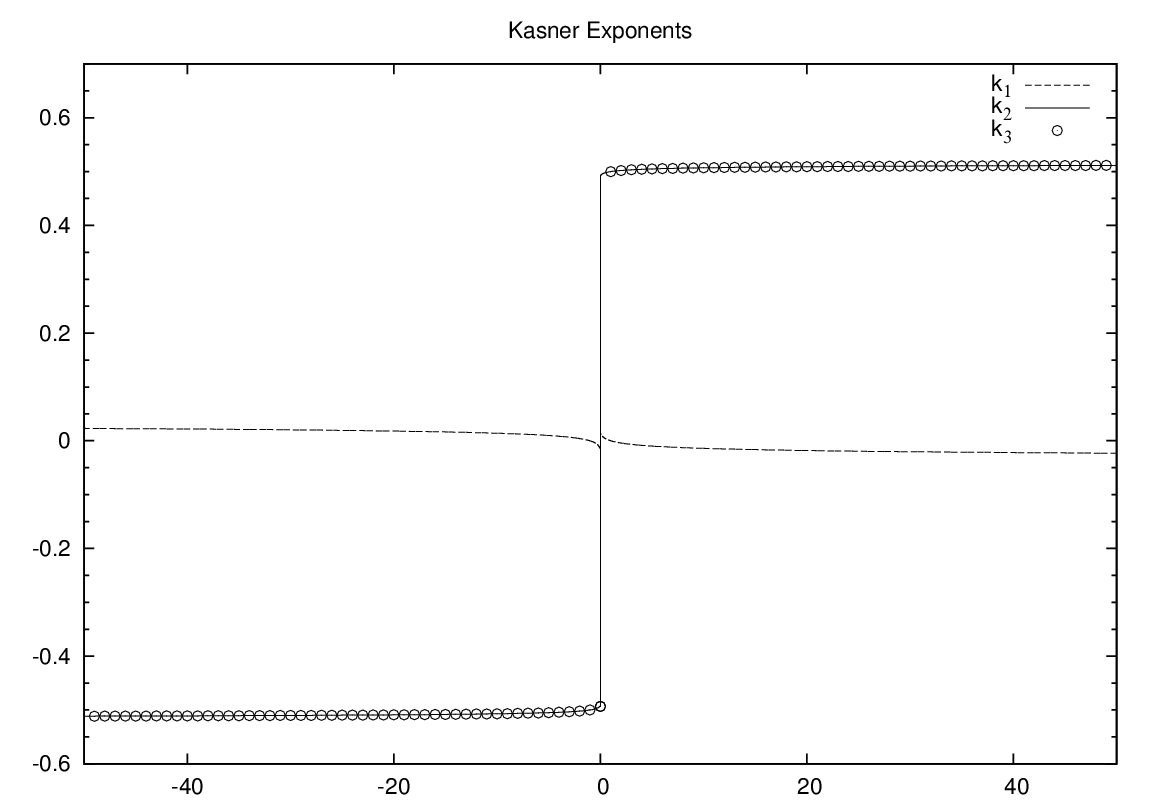}\\
\includegraphics[width=8cm]{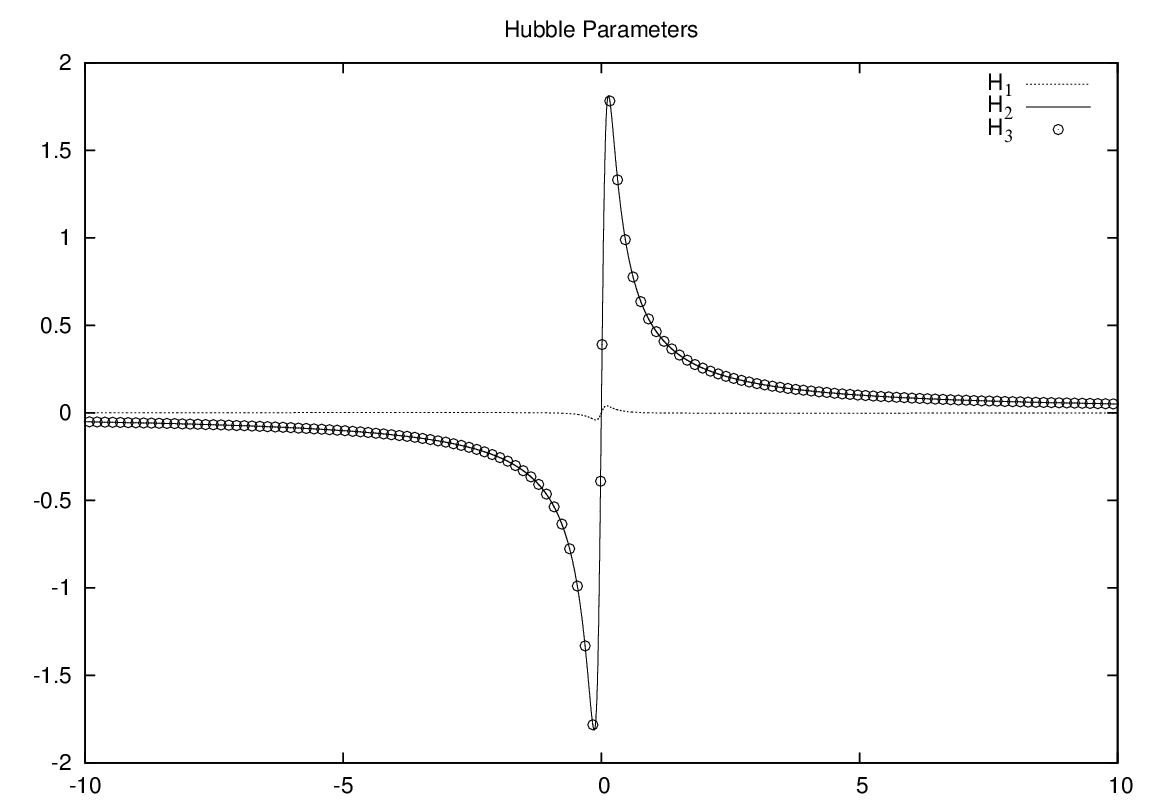}&
\includegraphics[width=8cm]{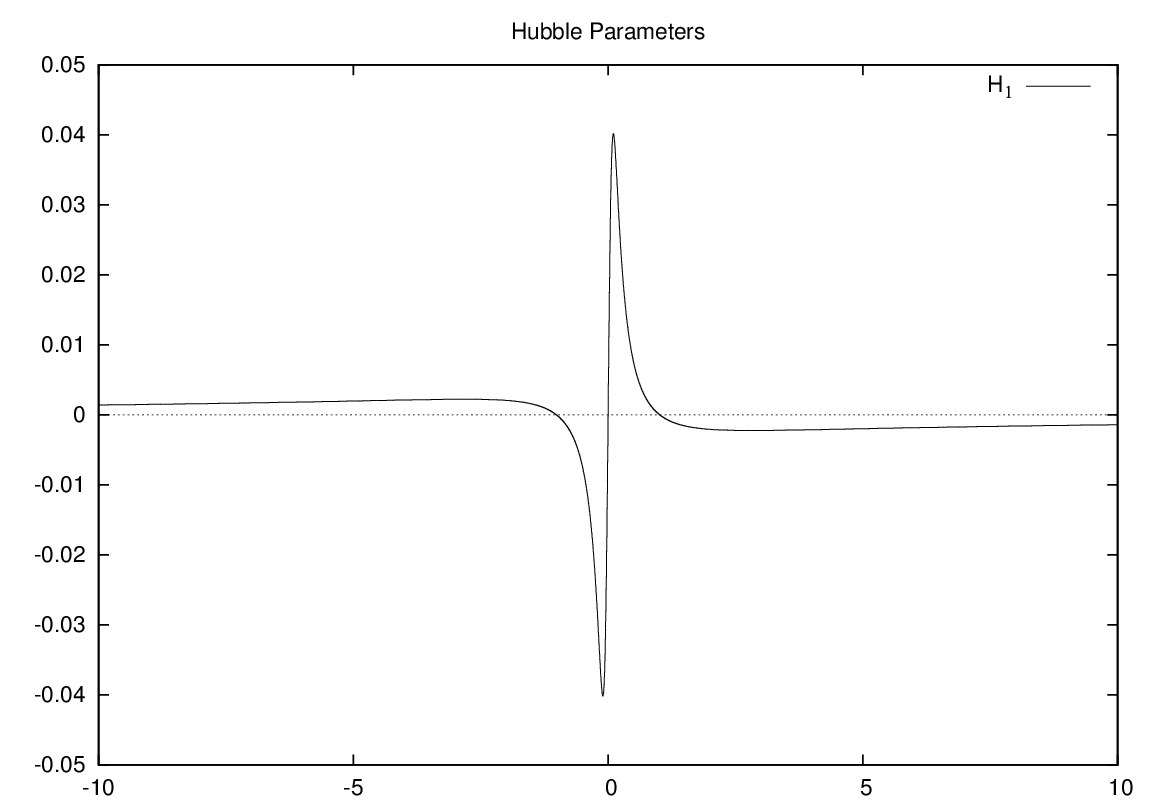}\\
\includegraphics[width=8cm]{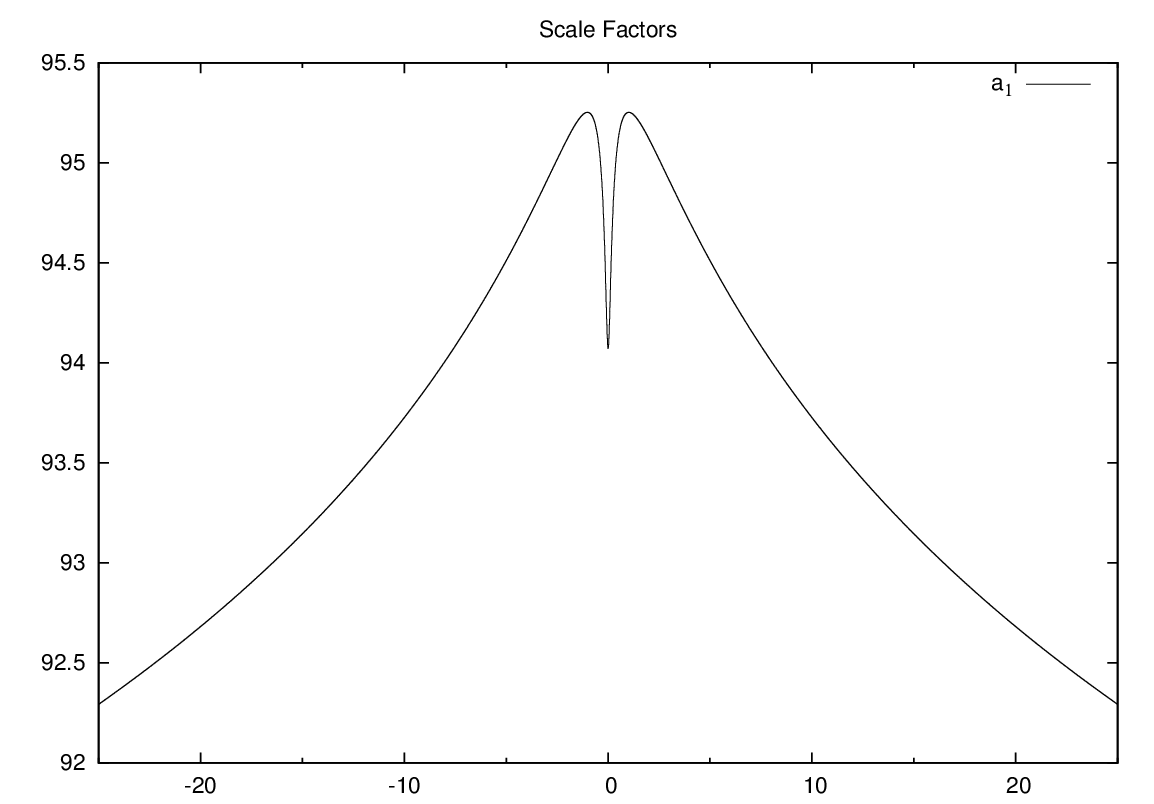}&
\includegraphics[width=8cm]{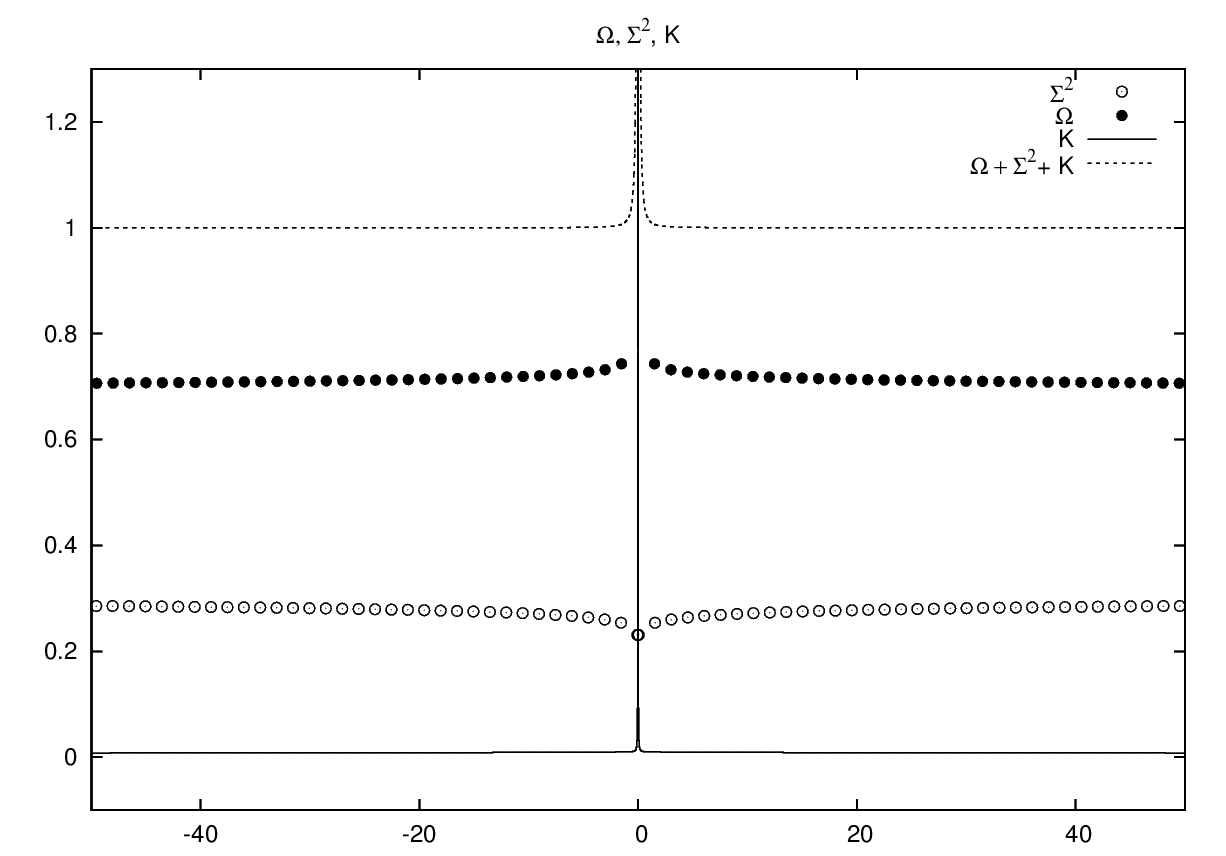}
\end{tabular}
\caption{LRS solution. The density reaches its maximal value when the shear is zero at the bounce, but the shear is different from zero in the rest 
of the evolution; the dynamical observables $a_i,H_i,k_i$ for directions $2$ and $3$ are equal, this shows that it is a LRS solution; 
the directional scale factor $a_1$ bounces one time and has two turn-around points; the curvature parameter $K$ is different from zero and asymptotically it satisfies  $\Omega+\Sigma^2 + K =1$.}
\label{fig:LRS}
\end{center}
\end{figure}

These subclass of models are characterized by the directional scalar factors $a_1, a_2, a_3$ such that $a_2=a_3$. 
In particular, this means that at each point the space-time is invariant under rotation about a preferred spacelike axis. 
In our variables this is written as $p_2=p_3$ and $c_2=c_3$ (if these equalities are satisfied by the initial conditions then 
they are satisfied through the evolution). 
These solutions will be useful to study the limit when the density is equal to the maximal density. In Bianchi II 
one might expect a maximal density strictly less than $1.315 \rho_{\rm crit}$ in all the solutions due to the presence of anisotropies. Here we show
that this is not the case, and the density can achieve the maximal value as a consequence of the shear being zero at the bounce. 
This is very different from Bianchi I models
where the presence of anisotropies makes the shear at the bounce always greater than zero. 

In order to have control on the density to make it maximal we put the initial conditions at the bounce, i.e., when 
$\bar\mu_1c_1=\bar\mu_2c_2=\bar\mu_3c_3= \pi/2$ and then evolve back and forward in time. Also, we need to make $x$
equal to the value that makes the density maximal, namely $x=\f{2}{(1+\gamma^2)\lambda}\approx 0.83$. If we take $p_2=p_3=1000$ (LRS condition)
then $p_1=113$ and using the fact that we are starting at the bounce ($\bar\mu_ic_i=\pi/2$), we calculate 
$c_1=64.98, c_2= c_3 = 7.34$ and, from the Hamiltonian constraint, $p_\phi = 11029.97$. 

The solution is plotted in figure \ref{fig:LRS}, where it is shown that in fact the solution 
has a maximal density at the bounce. The shear is zero at the bounce and has a non trivial behavior, such as two maxima.
In the evolution of the Kasner exponents we can see  (looking from left to right)  that two directions ($a_2, a_3$) are contracting and one 
direction ($a_1$) is expanding, and the 
Kasner exponents look like Bianchi I exponents far away from the bounce. 
This is because classically Bianchi II approaches Bianchi I when time goes to infinity \cite{wain_ellis,w_hsu,mac}, 
then in the classical region the Kasner exponents of LRS Bianchi II have as a limit the Kasner exponents of LRS Bianchi I. 
In figure \ref{fig:LRS} we plot the Hubble parameter $H_1$, which is the one that presents a new behavior. 
From the plot it can be seen that the direction $a_1$ 
has a bounce and two turn around points, which indicates also a new behavior that was not present in the isotropic and Bianchi I cases. 
The other two directions ($a_2,a_3$) bounces just one time for a total of three directional bounces and one global bounce ($\theta=0$).

There is a valid question at this point. Why can  the density reach its maximal value in Bianchi II and not in Bianchi I when there are anisotropies?
This question arises because in both models there are new degrees of freedom due to the anisotropies, so in principle they will have a similar behavior
respect to the distribution of the energy in gravitational waves. But the new feature in Bianchi II --not present in Bianchi I-- is the nontrivial spatial curvature. This curvature
gives a new degree of freedom with respect to the possible ways the energy density can be distributed. Now, the dynamical contribution not only comes from the matter density and the shear
but also from the spatial curvature. This fact can be quantitatively understood from the fact that the curvature parameter $K$ is non zero in Bianchi II.
As we can see in figure \ref{fig:LRS}, the plot for ($\Omega,\Sigma^2,K$) shows that the curvature parameter $K$ is different from zero. Thus, this provides also a qualitative explanation for the important difference between these two models, namely that the density can reach its maximal value at the bounce. This behavior must also occur in the Bianchi IX case \cite{CKM} where there is 
spatial curvature as well.

\subsection{Vacuum Limit} \label{sec:4:BII}
\begin{figure}[tbp!]
\begin{center}
\begin{tabular}{ll}
\includegraphics[width=8cm]{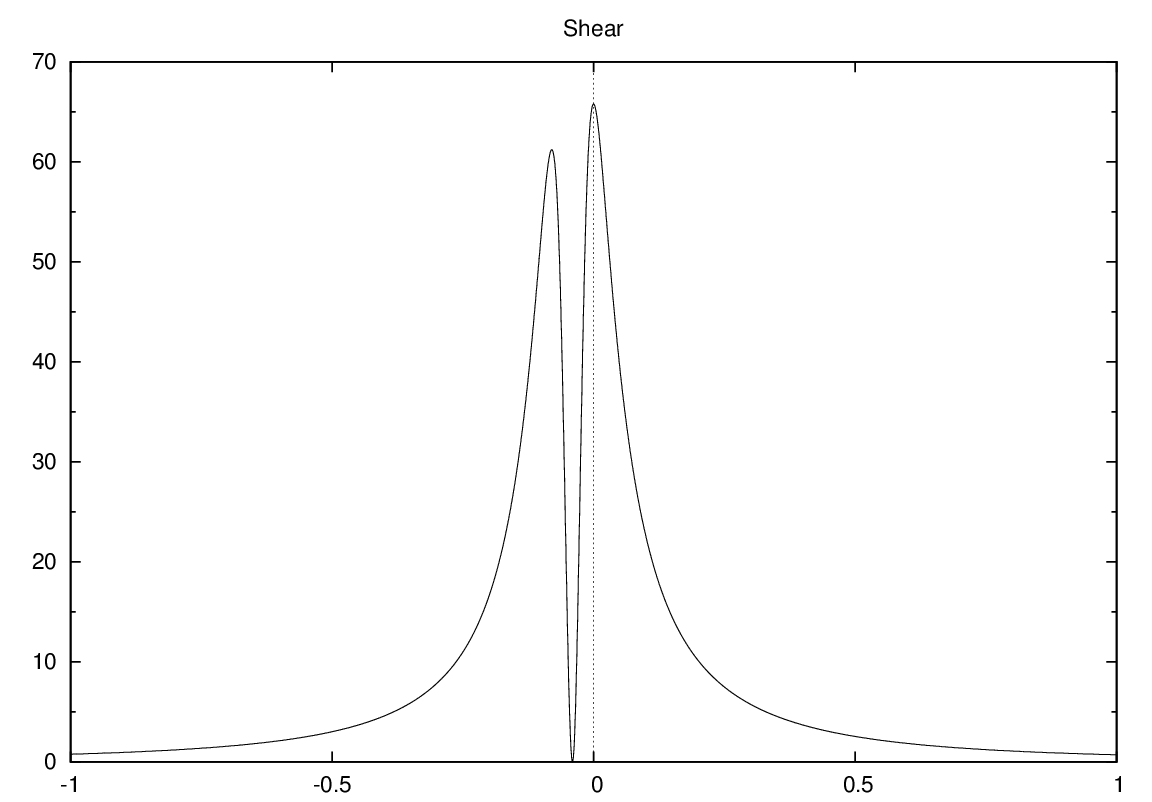}&
\includegraphics[width=8cm]{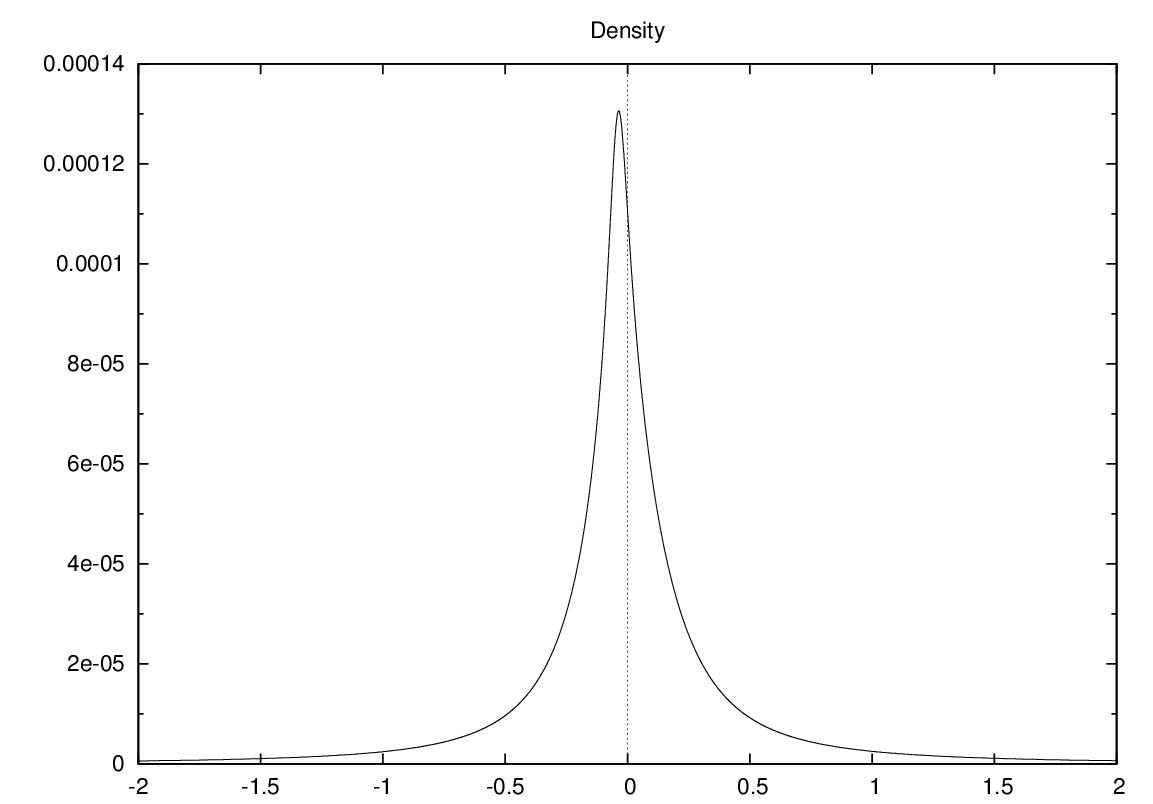}\\
\includegraphics[width=8cm]{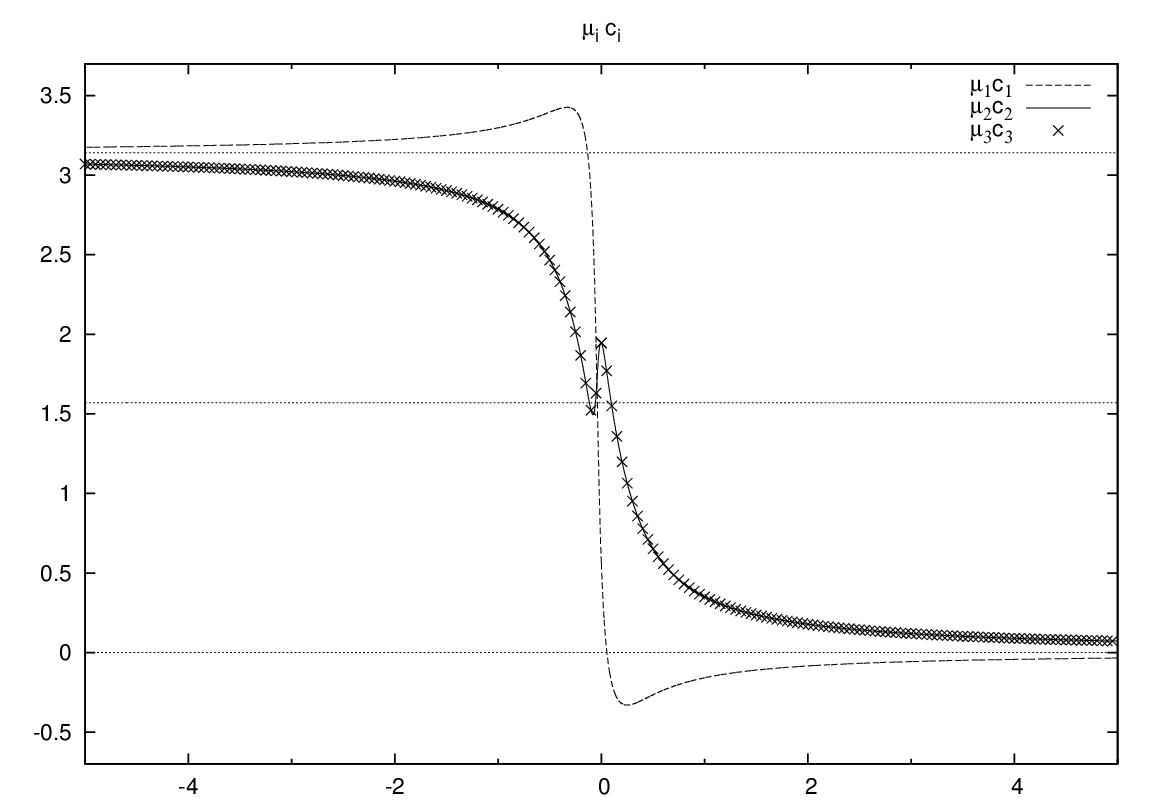}&
\includegraphics[width=8cm]{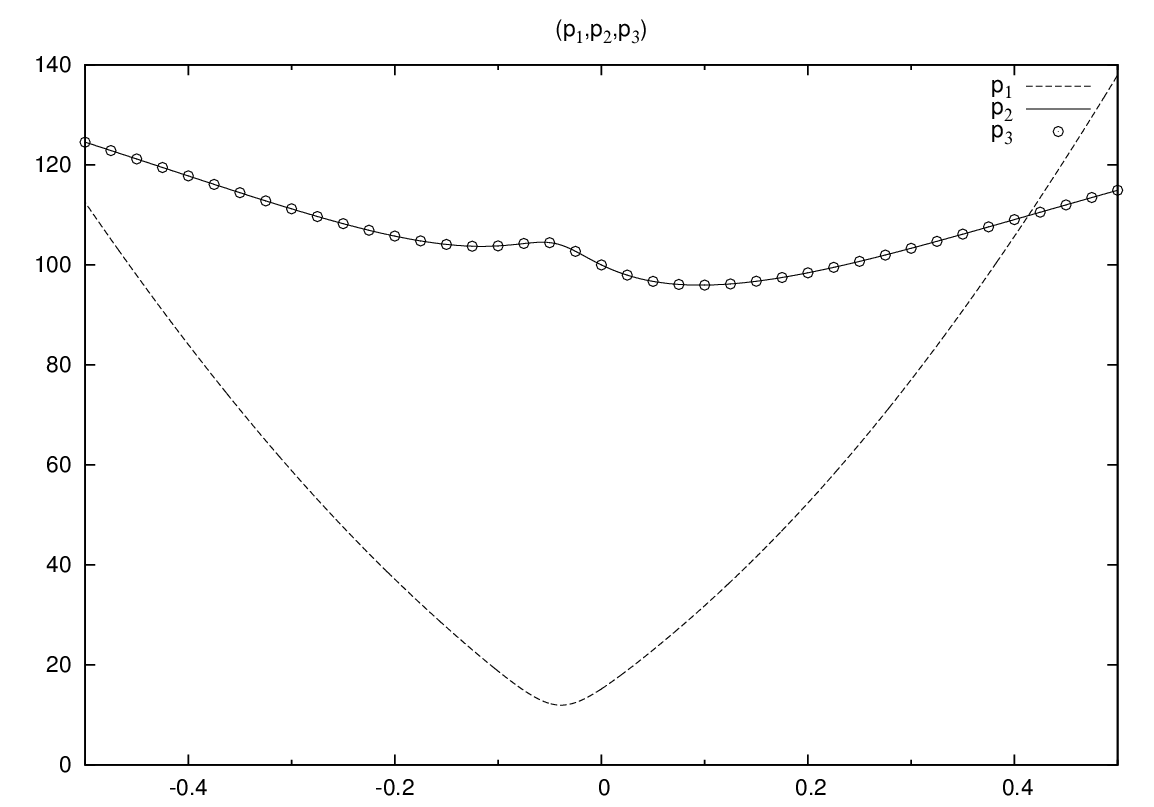}\\
\includegraphics[width=8cm]{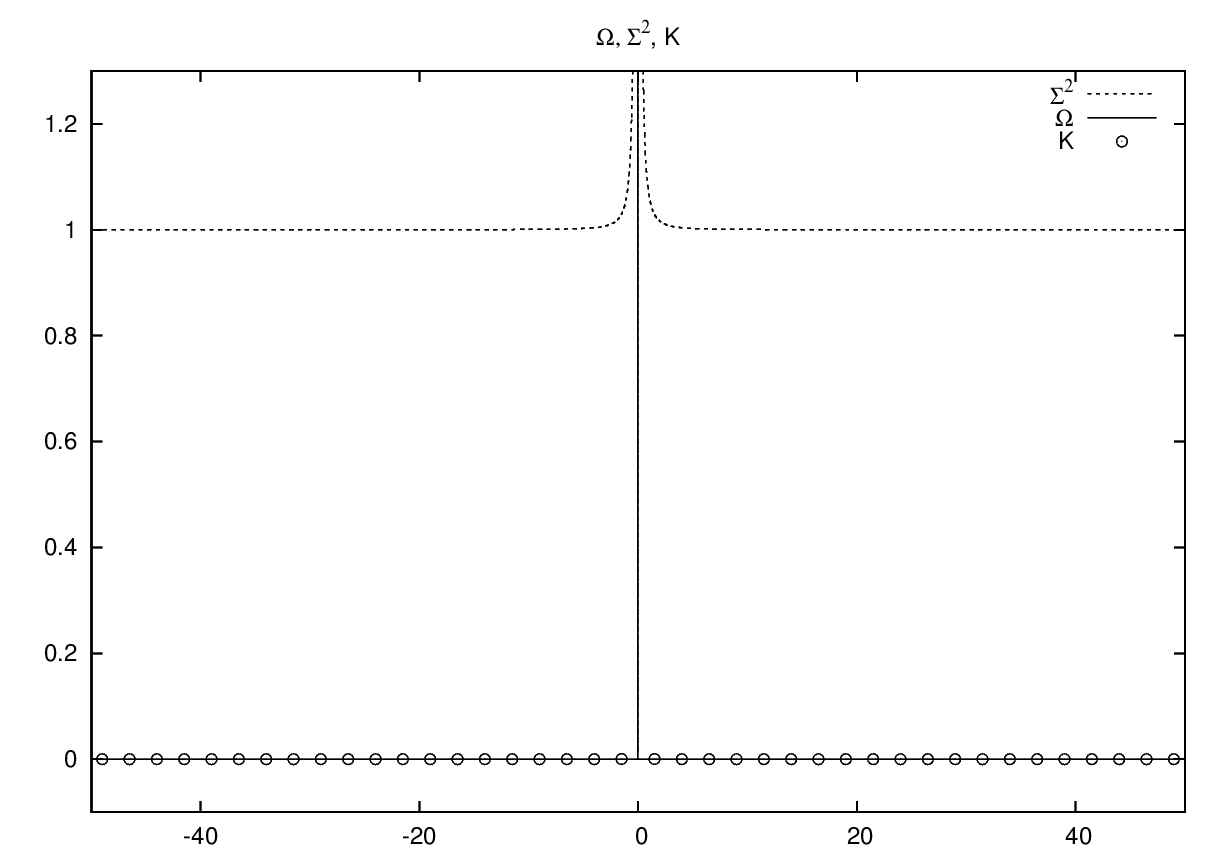}&
\includegraphics[width=8cm]{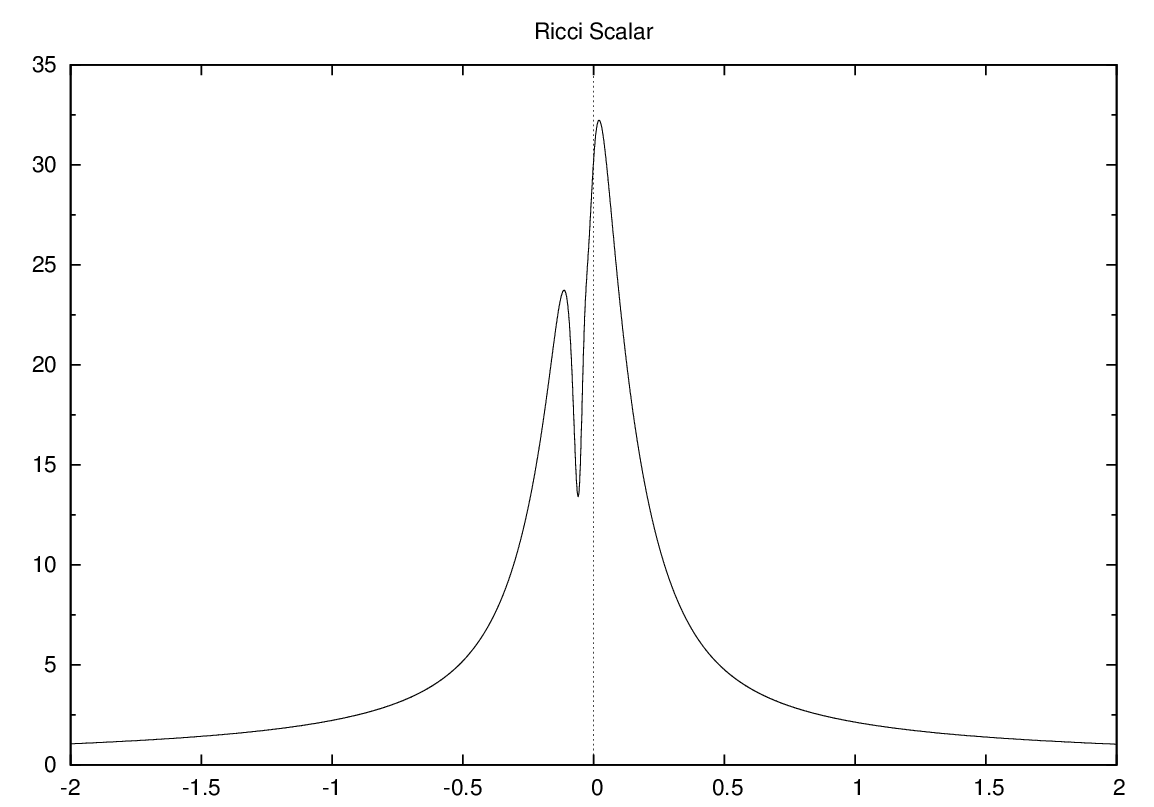}
\end{tabular}
\caption{Vacuum limit. The shear reaches its maximal value at one side of the bounce and it is zero on the other side (the zero is not at the bounce); 
the density is not zero but has a small value (even at the bounce); $\bar\mu_ic_i$ 
evolve from one classical solution ($\bar\mu_ic_i\approx 0$) to other one ($\bar\mu_ic_i\approx \pi$);
$\dot p_2$ and $\dot p_3$ are zero three times; all the dynamical contributions come from the shear ($\Omega\approx 0, K \approx 0, \Sigma^2\approx 1$);
the Ricci scalar reaches its maximal value at one side of the bounce and a local minimum on the other side. 
}
\label{fig:vacuum}
\end{center}
\end{figure}

Finally we study the solutions in the vacuum limit with maximal shear. As in the previous limiting cases, we do not want to study the vacuum 
case ($\rho=0$), but rather we want to study it as a limit approaching Bianchi II with a massless scalar field. That is, we shall study the solutions where
$\rho\rightarrow 0$ (or equivalently $p_\phi \rightarrow 0$). The problem is that
the density goes to zero when the time goes to $\pm\infty$  in all these solutions. 
Then we need to clarify what the vacuum limit means. In order to do this, we need to have in mind that the density has a 
maximal at the bounce, then we are interested in taking the vacuum limit $\rho\rightarrow 0$ in a finite time (near to the bounce). That is, we want solutions with density  near zero at the bounce or, equivalently, solutions where $\Omega \approx 0$ asymptotically. 
Among these class of solutions we select those with maximal shear, 
this allows us to study the extreme solutions where all the dynamical contribution comes from the anisotropies.

To obtain the initial conditions we found numerically the values of $x$ and $-\pi/2<\bar\mu_1c_1,\bar\mu_2c_2,\bar\mu_3c_3<3\pi/2$ 
that makes the shear maximal with a density near to zero. To do this we use the analytical expression for the shear found in \cite{singh} and the 
Hamiltonian constraint Eq. (\ref{effective-H}). The values found are: 
$x=1.69, \bar\mu_1c_1 = 0.5616,  \bar\mu_2c_2 = \bar\mu_3c_3 = 1.94618$. These values fix $4$ initial conditions, namely $p_1,c_1,c_2,c_3$,
as functions of the other two parameters that are (in some sense) not relevant for the behavior that we want to study. In our simulations, we take $p_2=100$ and $p_3=100$. 
The initial time is at $t=0$ (vertical line in the plots). 
The solution is shown in figure \ref{fig:vacuum}. The first unexpected behavior is that the shear is zero at the bounce and reaches its maximal value
on one side of the bounce. Moreover, $p_2$ and $p_3$ have a new behavior because 
$\dot p_2$ and $\dot p_3$ are zero three times, these zeros can be seen in the plot for $\bar\mu_ic_i$, 
where $\bar\mu_2c_2$ (and $\bar\mu_3c_3$) cross the $\pi/2$ line three times. This behavior is not present in Bianchi I nor 
in the solutions to Bianchi II studied in the previous sections. 
From the density plot we can see that it has a small value at the bounce and from the plot for the parameters ($\Omega,\Sigma^2, K$), it  can be appreciated that all the dynamical
contribution comes from the anisotropies, as was expected.

\section{Discussion}
\label{sec:6}

In this paper, we analyzed the numerical solutions of the effective equations that 
come from the improved LQC dynamics of the Bianchi II model  \cite{bianchiII}. We choose a massless
scalar field as the matter source. This effective theory comes from the construction of the full 
quantum theory and we expect that it gives some insights about the quantum dynamics of semiclassical 
states. The accuracy of the effective equations has been established in the isotropic cases 
and thus we expect that they should give an excellent approximation of the full quantum evolution for 
semiclassical states.

Let us summarize our results. We considered the Bianchi II case at the classical and effective level. As  
Bianchi I is a limiting case for Bianchi II, we use the previous results for Bianchi I in order to have 
control on our model. As we have seen, we recover Bianchi I as a limiting case when $x\rightarrow 0$ 
or equivalently $ p_1^3 \gg p_2p_3$. 
It is important to keep in mind that the Bianchi I model is a limiting case and is not contained within the Bianchi II model. 
The Bianchi I solutions are interesting by themselves, since
they give information about the asymptotic behavior of Bianchi II.

In order to determine how the classical singularities are resolved and how the effective equations evolve, we choose a set of observable quantities like density, shear, expansion, Ricci scalar, etc., and studied their evolution numerically. 
The equations of motion admit different limits that can 
be used to check the accuracy of the solutions and explore some new insights that the Bianchi II model offers. In order to systematically study 
these solutions, we started from the classical limit showing the way in which 
the effective solutions solve the singularities and reduce to the classical ones far away from the bounce. 
Next, we explored the isotropic limit included into
the Bianchi I limit when there are no anisotropies. We found that these solutions have a maximal density equal to the critical density $\rho_{\rm crit}$, the shear is close to zero
and presents a non trivial behavior because it has four maxima and
vanishes when the density bounces. 
The shear is not zero because this solution is not an exactly isotropic solution, but it is a solution very close to it. Recall that Bianchi I and, therefore, 
the isotropic solutions are not contained within the Bianchi II solutions. 
Later on,  we added anisotropies to the Bianchi I limit and showed  
that they reproduce the known solutions \cite{bianchiold}. Here we have shown that 
the classical cigar-like singularities are resolved like the point-like singularities, 
with singularity resolution understood in terms of geometrical observables being well behaved. We could not show numerically
the resolution of the barrel-like singularities because showing this implies a fine-tuning in the initial conditions, but we studied the 
limit of this kind of singularities, and there is nothing that indicates that they are not resolved, too.
Then, we considered the Locally Rotationally Symmetric (LRS) model of Bianchi II 
and explored how to find the solutions with maximal density at the bounce.
In this model at each point the space-time is invariant under rotation about a preferred spacelike axis.  Here we showed that the density can have the maximal value as a consequence of the shear being zero at the bounce. This shear has a non trivial behavior, such as two maxima.
The Kasner exponents look like Bianchi I exponents far away from the bounce, which is consistent with the fact that  classically Bianchi II approaches Bianchi I when proper time goes to infinity \cite{wain_ellis,w_hsu,mac}.
It was then shown that one directional scale factor ($a_1$) can change its behavior up to three times 
(it has one bounce and two turn-around points) and the other two directions 
($a_2,a_3$) bounce once, for a total of three directional bounces and one global bounce (when the expansion is zero).

We also studied the solutions in the vacuum limit with maximal shear. 
This allowed us to study the extreme solutions where all the dynamical contributions come from the anisotropies.
These solutions present a small value of the density at the bounce and 
unexpected shear and Ricci scalar behaviors because they are asymmetrical, reaching their maximal value on one side of the bounce.
We found that there are generic solutions in which 
the matter density is larger that its value in the isotropic solutions, with point-like and cigar-like singularities.
Some important solutions are LRS, such as the one with maximal density and the vacuum limit.

In order to have control over the solutions we found that the best way to do it is to put the initial conditions at the bounce (or near to it). 
There are also two important points that we have checked in this numerical work. The first one is the convergence of the solutions and the 
second one is the evolution of conserved quantities, additionally to the physical tests that the program must pass.

Finally, these results can be used as a starting point to study the Bianchi IX model in order to know if the approach to the singularity of the effective
solutions present Bianchi I behavior with Bianchi II transitions, as happens with the classical solutions in the BKL conjecture.  
In case of the full LQC dynamics it would be interesting to know whether the evolution of the semi-classical states reproduce all the new 
rich behavior that we get from the effective theory. From this point of view our work can be seen as the first step in this direction, since we already have 
a systematic study of the solutions of the effective theory. A similar study for the Bianchi IX model in underway \cite{CKM}.

\section*{Acknowledgments}

\noindent We thank A. Ashtekar, Ed. Wilson-Ewing and D. Sloan for helpful discussions and comments.
This work was in part supported by DGAPA-UNAM IN103610, by NSF
PHY0854743 and by the Eberly Research Funds of Penn State. 

\begin{appendix}

\section{Convergence and Conservation}
\label{app:a}

\begin{figure}[htbp!]
\begin{center}
\begin{tabular}{ll}
\includegraphics[width=8cm]{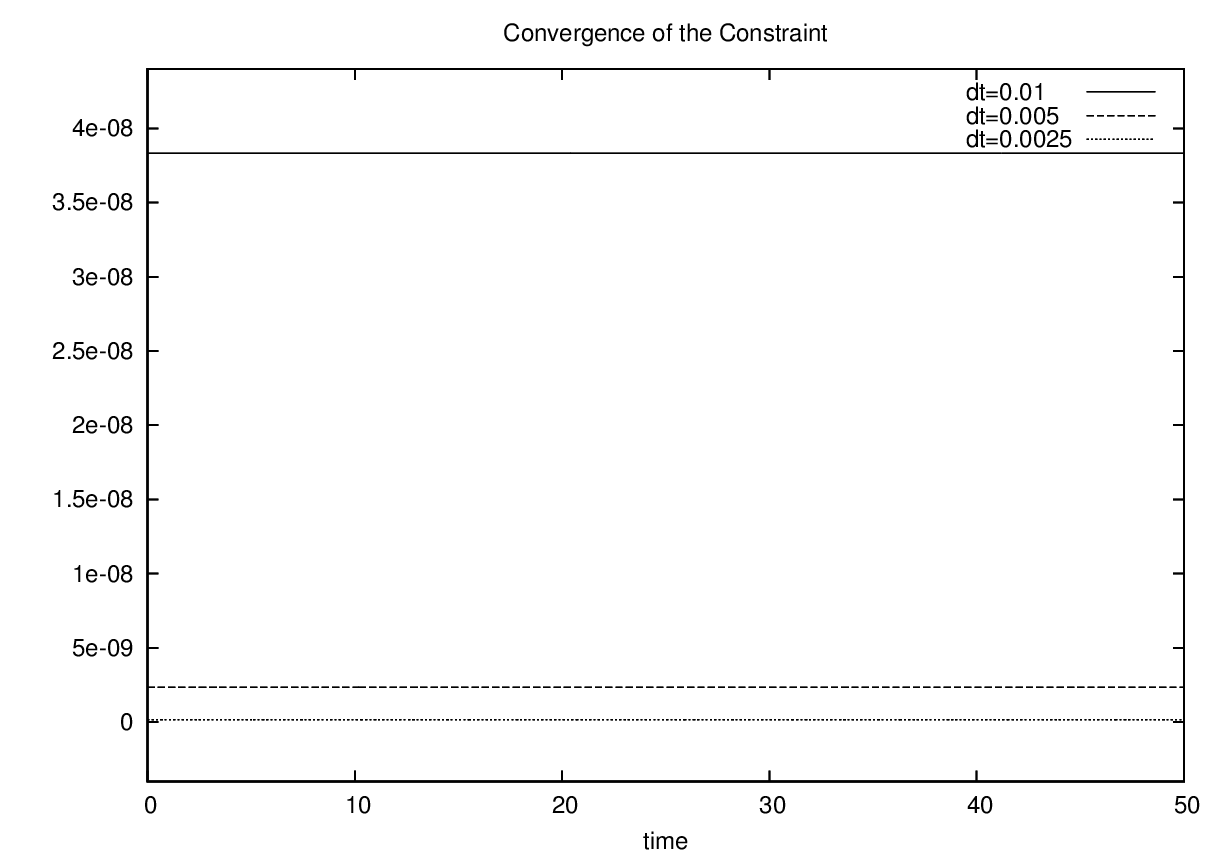}&
\includegraphics[width=8cm]{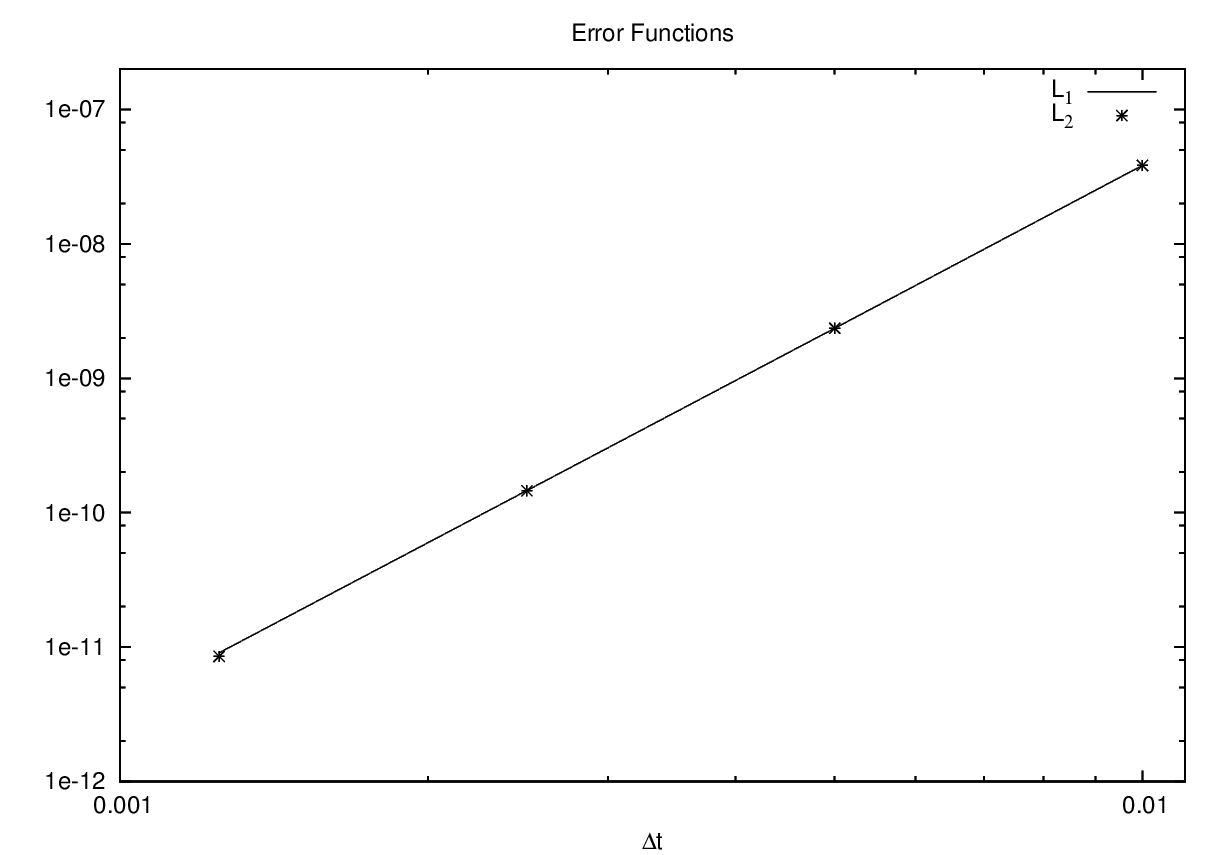}
\end{tabular}
\caption{Convergence of the constraint. The first plot shows the convergence of the solutions and that the constraint is conserved.
The second plot shows the error functions versus the bigger resultion used to calculate their.}
\label{fig:convergence}
\end{center}
\end{figure}

There are two important points that need to be cover when there is a numerical work: one is the convergence of solutions and other is
the evolution of conserved quantities. This means that, additionally to the physical tests that the program must pass, also must present a convergence 
of solutions, i.e., numerical solutions approach to analytical ones when the accuracy is improved, this say us that we are near
to the analytical solution with a small relative error. Remember that numerical solutions (in general) are never on the analytical solutions, 
all we can say is that they converge to them. Numerical solutions must also evolve on the constraint surface and they must preserve the 
conserved quantities,  this ensures that they are evolving on the physical phase space. In figure \ref{fig:convergence} it is shown the convergence 
of Hamiltonian constraint ($\mc{C}_H\approx 0$) when the time step is reduced, this also show that the constraint (or equivalently $p_\phi$) 
is conserved. The quantities plotted in figure \ref{fig:convergence} are the relative 
error for the constraint 
\be
\f{|(\mc{C}_H)_{\rm init}-\mc{C}_H(t)|}{(\mc{C}_H)_{\rm init}} \Leftrightarrow \f{|(p_\phi)_{\rm init}-p_\phi(t)|}{(p_\phi)_{\rm init}} \, ,
\ee
with different resolutions, and the error functions 
\be
L_1 = {\rm Max} |\mc{C}_H(t)_{\rm 2}- \mc{C}_H(t)_{\rm 1}|\, , \quad
L_2 = {\rm Max} \sqrt{|\mc{C}_H^2(t)_{\rm 2}- \mc{C}_H^2(t)_{\rm 1}|}\, ,
\ee
where the subindices in the Hamiltonian constraint mean resolution $2$ (with $\d t_2$) and resolution $1$  (with $\d t_1$), where 
$\d t_2=\d t_1/2$. The method used to integrate the equations is a Runge-Kutta 4 (RK4), while the resolutions used for the convergence tests are 
$dt = 0.01, 0.005, 0.0025, 0.00125, 0.000625$. The error functions for $(c_i,p_i)$ present similar behaviors. We can define the convergence order as
\be
n=\f{f_1-f_2}{f_2-f_3},
\ee
where $f_i$ is any evolved function at resolution $i$, with $dt_i>dt_{i+1}$. 
The convergence factor $n$ for a RK4 must be $n=2^4=16$. We obtain in our solutions $n\approx 16.2$, which
say us that solutions convergence as fast as expected.

\end{appendix}

\end{document}